\documentclass[letterpaper,12pt]{article}

\newcommand{\iid}{\stackrel{\mathrm{iid}}{\sim}}
\newcommand{\ind}{\stackrel{\mathrm{ind}}{\sim}}

\newcommand\independent{\protect\mathpalette{\protect\independenT}{\perp}}
\def\independenT#1#2{\mathrel{\rlap{$#1#2$}\mkern2mu{#1#2}}}

\usepackage{amsthm}
\usepackage{amsmath}
\usepackage{amssymb, amsfonts}
\usepackage{natbib}
\usepackage{bm}
\usepackage{graphicx}
\usepackage{color}
\usepackage{floatrow}
\usepackage{units}
\usepackage{url}
\usepackage{verbatim}
\usepackage{geometry}
\usepackage{slashbox}
\usepackage{epstopdf}
\usepackage{booktabs,caption,fixltx2e}
\usepackage[flushleft]{threeparttable}
\usepackage{rotating}
\usepackage{multirow}

\bibpunct{(}{)}{;}{a}{}{,}

\begin{document}
\title{Bayesian Nonparameteric Multiresolution Estimation for the American Community Survey}
\date{\today}
\author{Terrance D. Savitsky\thanks{U.S. Bureau of Labor Statistics, 2 Massachusetts Ave. N.E, Washington, D.C. 20212 USA}}

\maketitle
\begin{abstract}
Bayesian hierarchical methods implemented for small area estimation focus on reducing the noise variation in published government official statistics by borrowing information among dependent response values.  Even the most flexible models confine parameters defined at the finest scale to link to each data observation in a one-to-one construction.  We propose a Bayesian multiresolution formulation that utilizes an ensemble of observations at a variety of coarse scales in space and time to additively nest parameters we define at a finer scale, which serve as our focus for estimation.  Our construction is motivated by and applied to the estimation of $1-$ year period employment levels, indexed by county, from statistics published at coarser areal domains and multi-year intervals in the American Community Survey (ACS). We construct a nonparametric mixture of Gaussian processes as the prior on a set of regression coefficients of county-indexed latent functions over multiple survey years.  We evaluate a modified Dirichlet process prior that incorporates county-year predictors as the mixing measure.  Each county-year parameter of a latent function is estimated from multiple coarse scale observations in space and time to which it links.  The multiresolution formulation is evaluated on synthetic data and applied to the ACS.
\end{abstract}


\section{Introduction} \label{motivation}
The Local Area Unemployment Survey (LAUS) program of the U.S. Bureau of Labor Statistics (BLS) publishes employment and unemployment levels for all counties and municipal civil divisions (MCDs) (each of which nests within a county) across all states in the U.S.  The LAUS program uses by-county and MCD published employment statistics from the American Community Survey (ACS) to compute local allocation proportions of state employment levels.  The ACS is a national survey, conducted annually by the U.S. Census Bureau (Census), that replaces the information formerly published in the decennial census long-form.  The LAUS program apply these local allocation proportions to published by-state employment estimates from the Current Population Survey to render the local estimates of employment.

The ACS publishes sampling-weighted ``direct estimates" (which we denote with the term, 'statistics').   (Direct estimates weight the response value for each household in the sample back to the population from which it is drawn by using a sampling weight that is inversely proportional to its inclusion probability to compose a total or mean statistic for each domain and time period of interest.)  Employment statistics are published at $1-$, $3-$ and $5-$ year time intervals (which we denote as ``periods") for each of a wide variety of geographic domains.  The longer time periods enable the collection and pooling of more household samples to improve the estimation precision or coefficient of variation (CV); hence, each period statistic corresponds to a single time interval computed from the total sample collected during that period. In addition to pooling household observations across years into multi-period intervals, the ACS also aggregates counties into larger geographic domains, such as metropolitan or micropolitan areas, to achieve a larger sample size that allows publication of $1-$ year period statistics.  Census determines which periods and geographic domains to publish statistics in the ACS based on the supporting population size in each geographic domain in order to ensure an acceptable CV; for example, $1-$ year period statistics are published for all geographic domains with populations $> 65000$, while $3-$ year period statistics are provided for populations $> 20000$ and $5-$ year period statistics are otherwise provided.  A domain for which $1-$ year period statistics are published will also have published $3-$ and $5-$ year period statistics, while a domain for which $3-$ year period statistics are published will also have published $5-$ year period statistics.  Most counties and MCDs in the U.S. are relatively small, such that only $26\%$ of all \emph{counties} have published ACS $1-$ year period statistics.

In order to apply a consistent proportion-based allocation scheme across all counties and MCDs, the LAUS program is forced to use the $5-$ year period statistics, which are published annually.  While new sample observations are added to the $5-$ year published statistics with each year, the resulting pooled, multi-year interval statistic is lagged and possibly overly smoothed, which may result in a failure of the allocation proportion scheme to capture near-term changes in economic conditions, such as the recent Great Recession, which may dramatically alter the estimated proportions from one year to the next.  Our inferential goal in this paper is to develop a modeling approach that will utilize the published ACS statistics provided at these varied time periods and spatial domains to estimate latent, $1-$ year period values for all counties and MCDs, such that the LAUS program may employ these model-based $1-$ year period estimates to construct their local allocation proportions for all counties and MCDs in lieu of $5-$ year period ACS statistics.

Bayesian hierarchical modeling is extensively used in small area estimation applied to survey direct estimates published as official statistics by government agencies with the goal to reduce estimation uncertainty by borrowing information among parameters indexed by spatial area and often time period \citep{ghosh:1998}.  The use of hierarchical modeling facilitates the borrowing of estimation strength by shrinking all or some subset of domain-period parameters to a common mean. Those domain-periods with higher (known) variances (due to a relatively lower number of observations used to compose the published direct estimate) are shrunk to a greater extent towards the common value for the applicable subset of domains.

Even the most sophisticated small area modeling approaches, however, parameterize each regression mean to be linked one-to-one with an observed data point \citep{hawala:2012}.  These models may not be used to extract denoised, single year estimates for over $74\%$ of those counties and MCDs that don't have available $1-$ year period ACS statistics.  While the recent work of \citet{2014arXiv1405.7227B} appears to develop estimates for small domains from larger ones, they allocate or apportion larger domain estimates.  They don't attempt to estimate latent values for finer areas nested within coarser ones that are viewed to generate the observed coarse estimates.

We introduce a Bayesian approach that constructs parameters to be indexed on a fine scale and nest within one or more coarse-level observations in space and time.  Our approach employs multiple coarse-level observations, each of which provide some information about a fine-level parameter that nests within it.  We will see in the sequel that the parameters represent de-noised county-level employment levels and are constrained to sum to the mean of each ACS published data point of the domain and time period that nest the counties represented by the parameters.  There are often multiple $1-$ and/or $3-$ year period statistics published for these coarser spatial domains that may be used to provide some information about the counties which exhaust them.  

Our approach also leverages the nesting of years within (multi-year) periods; for example, we use the $2008 - 2012$ ACS publications, which will provide three, $3-$ year period statistics (e.g. $2008-2010$, $2009-2011$, $2010-2012$).  In the case where the ACS publishes $3-$ year period statistics for county ``A", the parameter defined for $2010$ in county A would link to (or nest within) all three statistics.

We employ a flexible nonparametric mixture approach for estimation of regression coefficients used to construct county-by-year parameters of each function, which allows the data to shrink estimated posterior distributions of the functions towards sub-group means.  This data-induced dimension reduction permits identification of the functions estimated from the coarser set of statistics that nest them.  We refer to our approach as a ``multiresolution" formulation because it utilizes observations defined at varied areal or time period resolutions for estimation of the by-county functions.

We specify the parameterization for our multiresolution likelihood and construct our associated nonparametric model for estimating their parameters in Section~\ref{method}. A brief overview of our algorithm to sample the set of full conditional posterior distributions defined by our model is discussed in Section~\ref{mcmc}.  We present estimated results for the collection of county/MCD-year parameters from the ACS in Section~\ref{acs}.  We perform a simulation study to assess the accuracy of the ACS estimates in Section~\ref{sim} and offer a concluding discussion in Section~\ref{discussion}.

\section{Method} \label{method}
We begin exposition of our model formulation that will provide fine-scale, $1-$ year period employment estimates for all counties and MCD domains by introducing their parameterization and how they connect to the statistics published at coarser scales in a likelihood statement.  We will subsequently introduce the nonparametric prior distributions that specify our probability model.
\subsection{Multiresolution Parameterization}
\begin{figure}[!h]
\begin{center}
\includegraphics[width=4.0in, trim=0 10 0 10, clip]{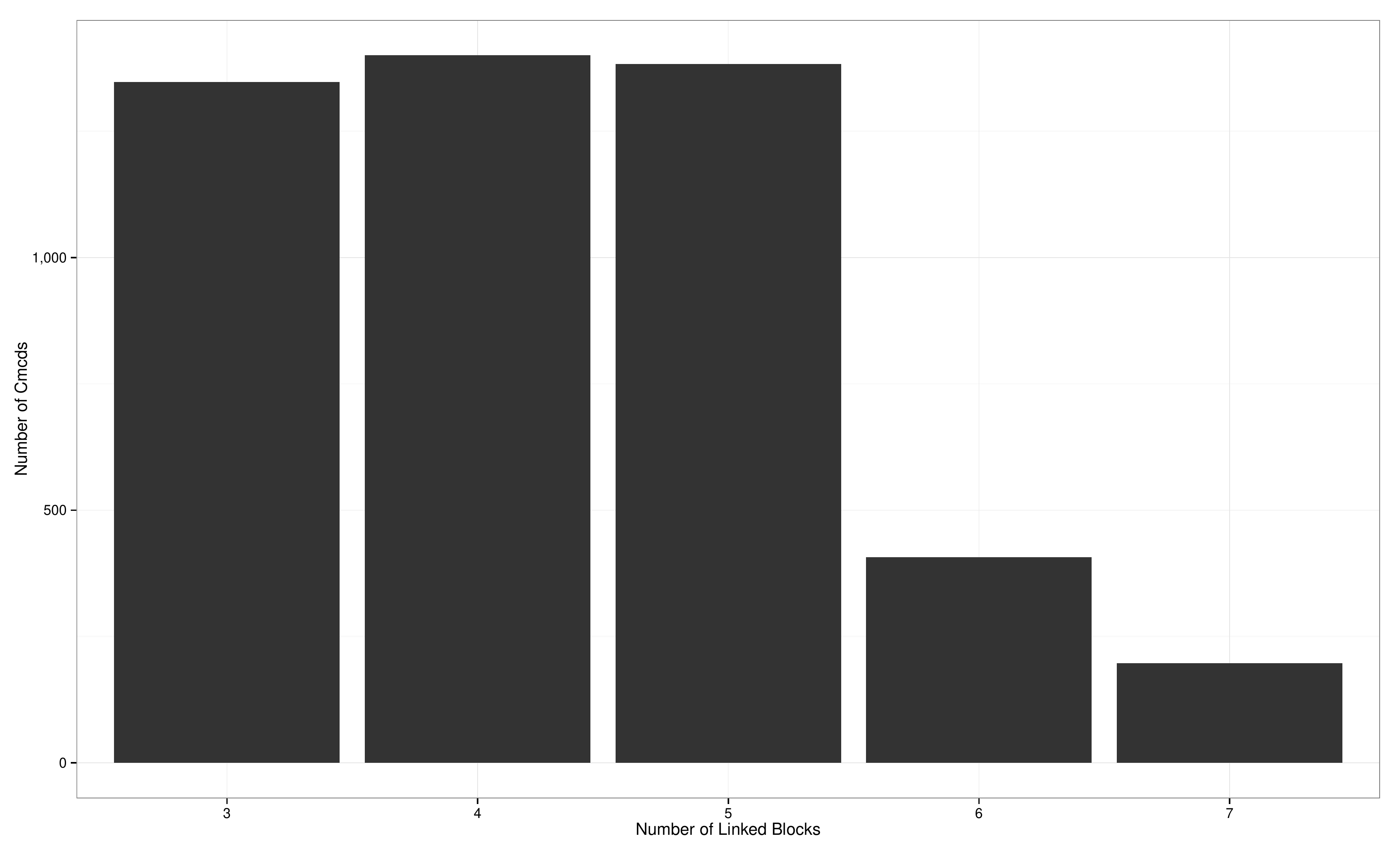}
\caption{Histogram of the number of block linkages for the $N = 4734$ counties.  The linkage counts includes the self-linkage.}
\label{countytoblocks}
\end{center}
\end{figure}

\begin{figure}[!h]
\begin{center}
\includegraphics[width=4.0in, trim=0 10 0 10, clip]{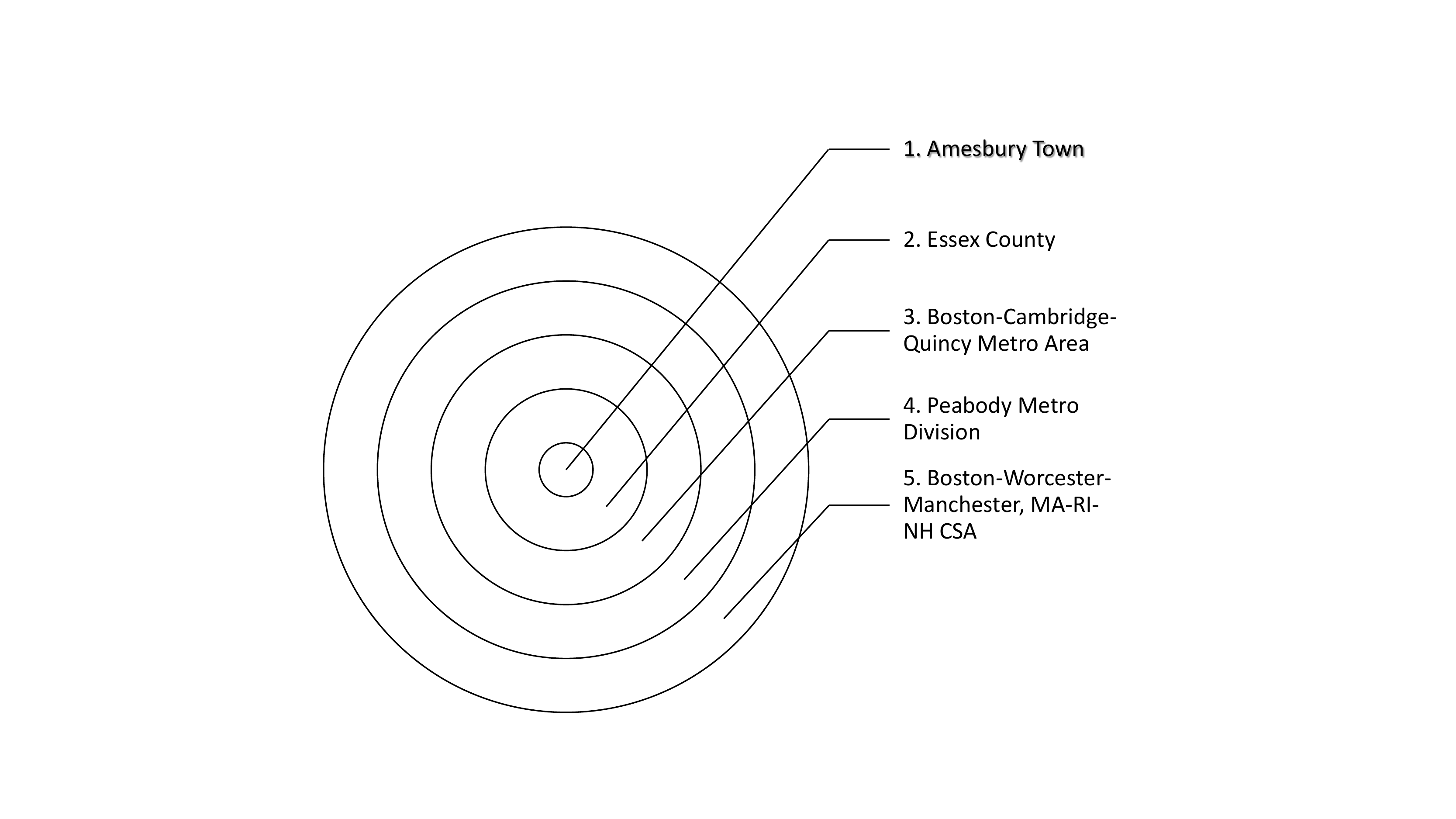}
\caption{Example of Block Nesting Structure for Amesbury Town, Massachusetts.}
\label{target}
\end{center}
\end{figure}

In the discussion to follow, we will use ``county" as a generic label to denote county and municipal civil division, the latter of which is primarily defined as a New England township designation where MCDs are nested within counties.  Let $f_{\ell j}$ denote the (latent) employment level for $\ell = 1,\ldots,(N = 4751)$ counties over years, $j = 2008,\ldots,2012$.  The counties are nested in larger core-based statistical areas (CBSAs), such as metropolitan (metro) and micropolitan (micro) areas, combinations of those larger areas (called core statistical areas or CSAs), including balance of states that subtract out all larger CBSAs and CSAs from each state.  Larger states generally have both metro and micro areas, as well as larger combinations of these.  (Census defines all CBSAs and CSAs to fully nest within a state).  Smaller states may have only one-to-a-few micro areas and no larger CSAs, other than the balance of state estimate that subtracts away the micro areas.   We denote all areas that geographically nest counties (which includes the counties, themselves) by the term ``block", $b = 1,\ldots,B$ and all counties nest in one or more blocks. We use published statistics for $B = 6074$ ACS blocks (that include the $N = 4751$ counties). Figure~\ref{countytoblocks} presents a distribution for the number of block links of the set of $N$ counties, from which we note that most counties link to $4-6$ blocks (including themselves).  Multiple block linkages occur because a county may nest within a block which is, in turn, nested within other blocks.  Figure~\ref{target} presents an example for Amesbury Town, Massachusetts, which links to $4$ other blocks through successive nestings.
We index the multi-year periods by $q = 1,\ldots,Q$, where each index value links a particular set of years.  Table~\ref{tab:periods} presents the set of years, $j$, (indexing the columns) that link with each period (row), $q$, where $1$ denotes a link and $0$, not.
\begin{table}[!h]
\begin{center}
{\footnotesize
\begin{tabular}{|c||l|l|l||l|l|l|}
  \hline
  \multirow{2}{*}{Period, $q$} & \multicolumn{5}{c|}{Year} \\
  \cline{2-6}
  & $2008$ & $2009$ & $2010$ & $2011$ & $2012$  \\
  \hline
  1 &   1 &   0 &   0 &   0 &   0 \\
  2 &   0 &   1 &   0 &   0 &   0 \\
  3 &   0 &   0 &   1 &   0 &   0 \\
  4 &   0 &   0 &   0 &   1 &   0 \\
  5 &   0 &   0 &   0 &   0 &   1 \\
  \hline
  6 &   1 &   1 &   1 &   0 &   0 \\
  7 &   0 &   1 &   1 &   1 &   0 \\
  8 &   0 &   0 &   1 &   1 &   1 \\
  \hline
  9 &   1 &   1 &   1 &   1 &   1 \\
  \hline
\end{tabular}}
\caption{Period, $q = 1,\ldots,(Q=9)$ links to years, $j = 2008,\ldots,2012$ \label{tab:periods}}
\end{center}
\end{table}

We may create a simple likelihood statement for each block-period statistic, $y_{bq}$, based on those counties, $\left(\ell\right)$, that nest in block, $b$ and those years, $\left(j\right)$, that nest in associated period, $q$, with,
\begin{eqnarray} \label{modellike}
y_{bq} &\ind& \mathcal{N}\left(\mathop{\sum}_{\ell \in b}\mathop{\sum}_{j \in q}f_{\ell j},\sigma^{2}_{bq}\right) \label{like}\\
f_{\ell j} &=& \mathbf{x}_{\ell j}^{'}\bm{\beta}_{\ell j},
\label{regression}
\end{eqnarray}
where the associated block-period variances, $\{\sigma^{2}_{bq}\}$, are \emph{known}. We observe that the $\left(f_{\ell j}\right)$ are constrained to sum to the de-noised mean of each observation, $y_{bq}$, which nests the associated counties and years. A $P\times 1$ county-year set of predictors, $\mathbf{x}_{\ell j}$, is incorporated into the model for the function, $f_{\ell j}$, with associated $P\times 1$ coefficients, $\bm{\beta}_{\ell j}$.  We construct $\mathbf{x}_{\ell j}$ with an intercept and a set of predictors defined at the county-year level available from administrative data.  The Quarterly Census of Employment and Wages (QCEW) is a census instrument targeted to business establishments (rather than households targeted by the ACS) that collects employment levels (on a monthly basis), which we aggregate to county and year. Our QCEW county-year predictors are employment levels for $12$ ``super sectors" defined in the North American Industry Classification System (NAICS): 1. Agricultural; 2. Natural resources and mining; 3. Construction; 4. Manufacturing; 5. Trade, transportation, utilities; 6. Information; 7. Financial activities; 8. Professional and business services; 9. Leisure and hospitality; 10. Other services; 11. Public Administration; 12. Unclassified.  We intend these $12$ predictors, together, to describe the composition of the economic activity for each county, by year, which we believe may provide a root-cause driver for employment level statistics.  We also include state records of unemployment claims aggregated to counties in our predictor set as a measure of economic health.  Our predictors will be critical to identify the regression coefficients and to regulate the borrowing of information for their estimation (through shrinkage).

We next define the prior distributions that permit flexibility in the borrowing of information for shrinkage in the estimation of the county-year regression coefficients.

\subsection{Prior on Functions}
The parametrization of Equation~\ref{regression} collects the $P \times T$ matrix of coefficients, $\mathbf{B}_{\ell} = \left(\bm{\beta}_{\ell 1},\ldots,\bm{\beta}_{\ell T}\right)$, indexed by county, $\ell = 1,\ldots,N$, on which we impose a conditional matrix variate Gaussian prior,
\begin{equation}
\mathbf{B}_{\ell} \ind \mathbf{0} + \mathcal{N}_{P\times T}\left(\mathbf{\Lambda}_{y,\ell}^{-1},\mathbf{C}\left(\bm{\kappa}_{\ell}\right)\right)\label{matprior},
\end{equation}
under the notation of \citet{dawid:1981}, where the $P \times P,~\mathbf{\Lambda}_{y,\ell}$ represents the precision matrix for the set of $P\times 1$ columns of $\mathbf{B}_{\ell}$ and the $T\times T,~\mathbf{C}(\bm{\kappa}_{\ell})$, denotes the covariance matrix for the rows of $\mathbf{B}_{\ell}$. The county-indexed covariance matrix, $\mathbf{C}_{\ell}$, is parameterized by $\bm{\kappa}_{\ell}$.  This specification is equivalent to the $TP \times TP$ covariance matrix constructed as $\mathbf{\Lambda}_{y,\ell}^{-1} \otimes \mathbf{C}(\bm{\kappa}_{\ell})$ under a multivariate Gaussian prior on the vector obtained by stacking the rows of $\mathbf{B}_{\ell}$.  The separable or tensor form we use for the covariance matrix reflects parsimony relative to a general $TP \times TP$ covariance matrix.  Yet, our parameterization for the latent functions is more flexible than that \citet{hawala:2012} who define $f_{\ell j} \sim \mathcal{N}(u_{\ell}+\mathbf{x}_{\ell j}^{'}\bm{\beta}_{j},\sigma^{2})$ (and each $f_{\ell j}$ is linked, one-to-one, to observation, $y_{\ell j}$, differently from our multiresolution construction, such that their model may not be employed to extract county-level, $1-$ year period estimates from the ACS).

We fix a particular county, $\ell$, and introduce the Gaussian process covariance formulation we construct for each of the $P,~T \times 1$ rows of $\mathbf{B}_{\ell} = \left(\bm{\beta}_{\ell 1},\ldots,\bm{\beta}_{\ell P}\right)^{'}$. The parameters, $\bm{\kappa}_{\ell}$, are used to specify a covariance formula for each cell of $\mathbf{C}(\bm{\kappa}_{\ell})$. Selecting (the $T \times 1$) row, $p$, of $\mathbf{B}_{\ell}$, the covariance formula is specified with,
\begin{eqnarray*}
\mathbf{C}\left(\bm{\kappa}_{\ell}\right) &\equiv& \mathbf{C}_{\ell} = \left(C_{\beta_{\ell pj},\beta_{\ell pk}}\right)_{j,k \in\left(2008,...,2012\right)}\\
C_{\beta_{\ell pj},\beta_{\ell pk}} &=& \frac{1}{\kappa_{\ell ,1}}\left(1 + \frac{\left(t_{ij}-t_{ik}\right)^{2}}{\kappa_{\ell ,2}\kappa_{\ell ,3}}\right)^{-\kappa_{\ell ,3}},
\end{eqnarray*}
where $\bm{\kappa}_{\ell} = \left(\kappa_{\ell,1},\kappa_{\ell,2},\kappa_{\ell,3}\right)$, which parameterizes a rational quadratic covariance formula.  The rational quadratic covariance formula may be derived as a scale mixture (over $\kappa$) of more commonly-used squared exponential kernels, $1/\kappa_{1}\exp\left((t_{j}-t_{\ell})^{2}/\kappa\right)$ \citep{rasm:2006}.  The vertical magnitude of surfaces rendered from a GP with the rational quadratic covariance formula is directly controlled by $\kappa_{\ell,1}$, while $\kappa_{\ell,2}$ controls the mean length scale or period, and $\kappa_{\ell,3}$ controls smooth deviations from the mean length scale.  Our choice of the rational quadratic covariance formula is intended as a parsimonious specification for parameterizing the use of a single covariance matrix, rather than utilizing a sum or product of multiple covariance matrices, each under the simpler squared exponential covariance formula.  See \citet{savitsky2011} for more background on the Gaussian process covariance formulations.
Our GP prior, parameterized by the $T \times T$ covariance matrix, $\mathbf{C}(\bm{\kappa}_{\ell})$, under a rational quadratic formulation produces rows of $\mathbf{B}_{\ell}$ that are infinitely smooth (because they are differentiable at all orders), which will in turn, produce a smooth estimation for the $T \times 1$ de-noised function, $\mathbf{f}_{\ell}$.  The smoothness restriction helps separate signal captured in $\mathbf{f}_{\ell}$ from the rough, non-differentiable noise in the observations, $\left(y_{bq}\right)$, to which $\mathbf{f}_{\ell}$ is linked.  We believe this smoothness assumption is reasonable to separate signal from noise present in the ACS statistics and rely on it to help identify the regression coefficients.  The $P\times P$ precision matrix, $\mathbf{\Lambda}_{y,\ell}$, allows the data to estimate a dependence among the $P$ sets of $T \times 1$ functions, each drawn from the Gaussian process.

\subsection{Clustering the Distributions of the Coefficients, $\{\mathbf{B}_{\ell}\}$}\label{nopred}
Define $\mathbf{\Theta}_{\ell} = \{\mathbf{\Lambda}_{y,\ell}, \bm{\kappa}_{\ell}\}$, where we note that the indexing by county, $\ell = 1,\ldots,N$, in Equation~\ref{matprior} instantiates a marginal mixture (of matrix variate Gaussians) prior for $\left(\mathbf{B}_{1},\ldots,\mathbf{B}_{N}\right)$.  We will next define a non-parametric prior distribution for $\mathbf{\Theta}_{\ell}$ that will allow the data to estimate probabilistic clusters, such that those counties, $\{\ell\}$, whose $\left(\mathbf{\Theta}_{\ell}\right)$ are assigned to the same cluster will draw their coefficients, $\left(\mathbf{B}_{\ell}\right)$, from the same Gaussian mixture component.  We (probabilistically) cluster the parameters of the Gaussian prior that generates each $\mathbf{B}_{\ell}$, rather than directly clustering the set of $\left(\mathbf{B}_{\ell}\right)$, because we don't expect any of the coefficients (and associated $T \times 1$ functions, $\left(\mathbf{f}_{\ell}\right)$) to be exactly equal.  Rather, we expect subsets of functions to be ``similar", which we define as drawing their coefficients (assigned to same cluster) from the same Gaussian distribution.

We specify a Dirichlet process prior for $\left(\mathbf{\Theta}_{\ell}\right)$ in,
\begin{subequations}
\label{dpmix}
\begin{align}
\mathbf{\Theta}_{1},\ldots,\mathbf{\Theta}_{N}\Big|G &\sim G \\
G \Big|\alpha,G_{0} &\sim \mbox{DP}(\alpha, G_{0}),
\end{align}
\end{subequations}
where $\left(\mathbf{\Theta}_{\ell}\right)_{\ell=1,\ldots,N}$ receive a random distribution prior, $G$, drawn from a Dirichlet process (DP), parameterized with a concentration parameter, $\alpha$, a precision parameter that controls the amount of variation in $G$ around prior mean, $G_{0}$. The base or mean distribution, $G_{0} = \mathcal{W}\left(P+1,\mathbb{I}_{P}\right) \times \mathop{\prod}_{d =1}^{D=3}\mathcal{G}a\left(a,b\right)$, a $P-$ dimensional Wishart distribution for the $P\times P,~\mathbf{\Lambda}_{y,\ell}$, and a product of Gamma priors for the $D = 3$ parameters in the rational quadratic specification for the parameters, $\bm{\kappa}$, that parameterize the $T \times T$ covariance matrix, $\mathbf{C}$, respectively.  Equation~\ref{dpmix} describes a mixture model of the form, $\mathbf{B}|G \iid \int \mathbf{0} + \mathcal{N}_{P\times T}\left(\mathbf{\Lambda}_{y},\mathbf{C}\left(\bm{\kappa}\right)\right)
G\left(d(\mathbf{\Lambda}_{y},\bm{\kappa})\right)$, where $G$ is the mixing measure over the precision and covariance parameters, $\mathbf{\Theta} = \{\mathbf{\Lambda}_{y}, \bm{\kappa}\}$.

The DP formulation may be described as approximating any unknown distribution by placing spikes at ``location" values in the support of $G$, which are each drawn from $G_{0}$, with heights equal to probability mass values associated to the locations, such that draws from $G$ are almost surely discrete.  The discrete construction for $G$ allows for ties among the $\left(\mathbf{\Theta}_{\ell}\right)$ that we interpret as probabilistic clusters.  We examine this clustering property of the DP by expressing it in the (stick breaking) form as a set of weighted locations \citep{sethuraman:1994},
\begin{equation}\label{stick}
G = \mathop{\sum}_{h=1}^{\infty} p_{h}\delta_{\mathbf{\Theta}^{\ast}_{h}},
\end{equation}
where $G$ is a countably infinite mixture of weighted point masses with ``locations", $\mathbf{\Theta}^{\ast}_{1},\ldots,\mathbf{\Theta}^{\ast}_{M}$, indexing the unique values for the $\left(\mathbf{\Theta}_{\ell}\right)$, where $M \leq N$ (counties from the finite population).  We record cluster memberships of counties with $\mathbf{s} = \left(s_{1},\ldots,s_{N}\right)$ where $s_{\ell} = \ell$ denotes $\mathbf{\Theta}_{\ell} = \mathbf{\Theta}^{\ast}_{\ell}$ so that $\{\mathbf{s},\left(\mathbf{\Theta}^{\ast}_{m}\right)\}$ provides an equivalent parameterization to $\left(\mathbf{\Theta}_{\ell}\right)$ and we recover $\mathbf{\Theta}_{\ell} = \mathbf{\Theta}^{\ast}_{s_{\ell}}$.  The weight, $p_{h} \in (0,1)$ is composed as $p_{h} = v_{h}\mathop{\prod}_{k = 1}^{h-1}\left(1-v_{k}\right)$ where $v_{h}$ is drawn from the beta distribution, $\mathcal{B}e\left(1,\alpha\right)$.  This construction provides a prior penalty on the number of mixture components, but we also see that a higher value for $\alpha$ will produce more clusters (unique locations).  Since each location is drawn from $G_{0}$, as the number of unique locations increases, the estimated $G$ approaches the base distribution, $G_{0}$.  We place a further gamma prior on $\alpha$ to allow posterior updating in recognition of the relatively strong influence it conveys on the number of clusters formed \citep{escobar:1995}.

\subsection{Predictor-Assisted Clustering}\label{predassist}
We have, so far, specified a likelihood linking subsets of county-year functions, $\left(f_{\ell j}\right)$, to each of the block-period statistics, $y_{bq}$.  The structure in our model is defined through the regression model on $f_{\ell j} = \mathbf{x}_{\ell j}^{'}\bm{\beta}_{\ell j}$, under the subsequent hierarchical prior formulation we constructed for $\left(\mathbf{B}_{\ell}\right)$.  If we had imposed the DP prior directly on the $\left(\mathbf{B}_{\ell}\right)$, the estimated functions would have been locally linear (for each subset of county-indexed coefficients assigned to same cluster), but globally non-linear.  We defined a nonparametric mixture prior for $\left(\mathbf{B}_{\ell}\right)$ by placing the DP prior on the covariance parameters, $\mathbf{\Theta}_{\ell} = \{\mathbf{\Lambda}_{y,\ell}, \bm{\kappa}_{\ell}\}$, of the Gaussian prior of Equation~\ref{matprior} such that the estimated functions will be both locally and globally non-linear.

The clustering of the counties is determined from the conditional distribution for $\mathbf{Y}= (y_{bq})\vert \left(\mathbf{X}_{\ell}\right)_{\ell = 1,\ldots,N}$ since we fix the predictors, $\left(\mathbf{X}_{\ell}\right)$.  Our estimation task is challenging because we will not have a one-to-one relationship between most block-period observations, $\mathbf{Y}$, and latent county-year parameters, $\left(f_{\ell j}\right)$.  So we would like to borrow the maximum amount of information provided in our data by incorporating the predictor values into the computation of probabilities for the co-clustering of the county covariance parameters of  $\{\mathbf{B}_{\ell}\}$.  If the $P\times T$ matrix of predictors, $\mathbf{X}_{\ell}$, for county, $\ell$, is very similar to, $\mathbf{X}_{\ell^{'}}$, for county, $\ell^{'}$, then we would like to define a higher prior probability for $\mathbf{\Theta}_{\ell} = \mathbf{\Theta}_{\ell}^{'} = \mathbf{\Theta}^{\ast}_{m}$, in which case $\mathbf{B}_{\ell}$ is drawn from the same matrix-variate Gaussian as $\mathbf{B}_{\ell^{'}}$, producing function $\mathbf{f}_{\ell}$ that is similar to $\mathbf{f}_{\ell^{'}}$.

We modify an approach of \citet{Muller:2011:PPM} to allow definition of a DP prior construction that incorporates the predictors, $\left(\mathbf{X}_{\ell}\right)_{\ell = 1,\ldots,N}$, into the determination of the clusters. We will treat the $P\times T$ predictor matrices, $\left(\mathbf{X}_{1},\ldots,\mathbf{X}_{N}\right)$, as though they were random (though we believe they are not random) as a computational device to induce the utilization of the predictors, as well as the response, in the estimation of the clustering (or partition) over county-indexed covariance parameters, $\left(\mathbf{\Theta}_{\ell}\right)$.  We next specify a probability model for the $\left(\mathbf{X}_{\ell}\right)$ and show how we will use it in determination of the cluster assignments,
\begin{subequations}
\label{xprior}
\begin{align}
\mathop{\mathbf{x}_{\ell j}}^{P\times 1} &\ind \mathbf{N}_{p}\left(\bm{\delta}_{\ell j}, \mathbf{H}_{x}^{-1}\right)\\
\mathop{\mathbf{\Delta}_{\ell}}^{P \times T} &\ind \mathbf{0} + \mathcal{N}_{P\times T}\left(\mathop{\mathbf{\Lambda}_{x,\ell}^{-1}}^{P\times P},\mathop{\mathbf{Q}(x,\ell)^{-1}}^{T\times T}\right)\\
\mathbf{Q}(x,\ell) &= \tau_{x,\ell}\left(\mathbf{D}_{x} - \rho_{x,\ell}\mathbf{\Omega}_{x,\ell}\right),
\end{align}
\end{subequations}
where $\mathbf{H}_{x} \sim \mathcal{W}\left(P+1,\mathbb{I}_{P}\right)$. $\mathbf{Q}(x,\ell)$ is constructed as a conditional autoregressive (CAR) prior \citep{rue:held:2005} that is similar in idea to the GP prior on $\mathbf{B}_{\ell}$, but tends to render rough, non-differentiable surfaces, rather than the smooth surfaces generated by a GP prior.  We use the CAR prior because it is computationally faster to draw posterior samples than the GP and we are not concerned with generating de-noised functions from $\mathbf{X}_{\ell}$, but only use the parameters of $\mathbf{X}_{\ell}$ to help determining the clustering of the covariance parameters of $\mathbf{B}_{\ell}$. The $T \times T,~\mathbf{D}_{x}$, is a diagonal matrix that sums the rows of the $T \times T,~\mathbf{\Omega}_{x}$, a similarity or adjacency matrix between pairs of time points (with zeros for the diagonal values). So each entry in $\mathbf{D}_{x}$ expresses the relative influence or precision for each time point.  The parameter, $\tau_{x,\ell}\sim\mathcal{G}a\left(a=1,b=1\right)$, controls the scale and, $\rho_{x,\ell}\sim\mathcal{U}\left(-1,1\right)$, controls the degree of autocorrelation.  The CAR prior may be heuristically thought of as a local, random walk smoother with a fixed length scale (unlike the GP, where the data estimate the length scale). See \citet{savitsky2013} for more details about the CAR prior.

We now extract $\{\mathbf{\Lambda}_{x,\ell}, \tau_{x,\ell}, \rho_{x,\ell}\}$ and simply expand $\mathbf{\Theta}_{\ell} = \{\mathbf{\Lambda}_{y,\ell}, \bm{\kappa}_{\ell}, \mathbf{\Lambda}_{x,\ell}, \tau_{x,\ell}, \rho_{x,\ell}\}$ under the DP prior of Equation~\ref{dpmix}, which now incorporates information about $\mathbf{X}_{\ell}$ into the clustering of $\mathbf{B}_{\ell}$.
To gain insight into how treating $\mathbf{X}_{\ell}$ as random influences the clustering mechanism, we present the kernel of the full conditional posterior distributions for the $N\times 1$ vector of cluster indicators, $\mathbf{s}$, after using the P\'{o}lya Urn scheme \citep{sethuraman:1994} to marginalize out the random measure, $G$,
\begin{equation}
\begin{split}
\quad f\left(s_{\ell} = s|\mathbf{s}_{-\ell},\mathbf{B}_{\ell},\mathbf{\Delta}_{\ell},\mathbf{\Theta}^{\ast}_s\right)\\
\quad \propto \frac{n_{s}-1}{N+\alpha}~\delta\left(s_{\ell} = s\right)L\left(\mathbf{B}_{\ell},\mathbf{\Delta}_{\ell}\right)\\
\quad + \frac{\alpha}{N + \alpha}~\delta\left(s_{\ell} = M^{-} + 1\right)L\left(\mathbf{B}_{\ell},\mathbf{\Delta}_{\ell}\right),
\end{split}
\label{polya}
\end{equation}
that is a product of the mixture prior, $f(s_{\ell}|\mathbf{s}_{-\ell}) = \frac{n_{s}-1}{N+\alpha}\delta\left(s_{\ell} = s\right) + \frac{\alpha}{N + \alpha}\delta\left(s_{\ell} = M^{-} + 1\right)$ (which assigns counties to clusters with probabilities proportional to their popularity, as measured by the number of counties assigned to cluster $s$, and with probability proportional to $\alpha$ generates a new cluster) and the joint likelihood,
\newline
$L\left(\mathbf{B}_{\ell},\mathbf{\Delta}_{\ell}\right) = \mathcal{N}_{P\times T}\left(\mathbf{B}_{\ell}\vert \mathbf{\Lambda}^{\ast}_{y,s},\mathbf{C}\left(\bm{\kappa}^{\ast}_{s}\right)\right)
\mathcal{N}_{P\times T}\left(\mathbf{\Delta}_{\ell}\vert \mathbf{\Lambda}^{\ast}_{x,s},\mathbf{Q}\left(\tau^{\ast}_{x,s},\rho^{\ast}_{x,s}\right)\right)$.
This computation reveals that the conditional posterior distribution for the cluster allocation of county $\ell$ is a function of \emph{both} the likelihood of $\mathbf{B}_{\ell}$, estimated from $\mathbf{Y} = \left(y_{bq}\right)$, and also that for $\mathbf{\Delta}_{\ell}$, which is estimated from $\mathbf{X}_{\ell}$. So the use of the joint likelihood in the full conditional posterior for the allocation of counties to clusters demonstrates that the cluster assignments are now controlled by the joint distribution for $\left(\mathbf{Y},\left(\mathbf{X}_{\ell}\right)\right)$.

\cite{Muller:2011:PPM} point out that is not necessary to believe the $\left(\mathbf{X}_{\ell}\right)$ are random in formulation of Equation~\ref{polya} that relies on Equation~\ref{xprior} to inject predictor information into the distribution over the clusterings (or partitions); rather, our assignment of a prior distribution to the $\left(\mathbf{X}_{\ell}\right)$, as part of a joint model with $\mathbf{Y}$, may be viewed as a computational device to implement a new prior distribution for the clusterings that incorporates the $\left(\mathbf{X}_{\ell}\right)$.  

The joint prior for the cluster indicators, $s_{1},\ldots,s_{N}$, under simpler model of Section~\ref{nopred} that parameterizes the conditional distribution for $\mathbf{Y}\vert \left(\mathbf{X}_{\ell}\right)_{\ell = 1,\ldots,N}$, is stated with,
\begin{equation}
f\left(s_{1},\ldots,s_{N}\right) \propto \alpha^{M-1}\mathop{\prod}_{m=1}^{M}(n_{m} -1)!, \label{basicprior}
\end{equation}
after marginalizing out the random measure, $G$, where, $n_{m} = \mathop{\sum}_{\ell = 1}^{N}\mathbb{I}\left(s_{\ell} = m\right)$ denotes the number of counties assigned to cluster, $m$.  As earlier noted, this prior for cluster assignments is independent of the predictor values, $\left(\mathbf{X}_{\ell}\right)$.

Our formulation that parameterizes a joint distribution for $\mathbf{Y},\left(\mathbf{X}_{\ell}\right)_{\ell = 1,\ldots,N}$ is equivalent to the model for $\mathbf{Y}\vert \left(\mathbf{X}_{\ell}\right)_{\ell = 1,\ldots,N}$, but with Equation~\ref{basicprior} adjusted to \emph{add} information about the predictors with,
\begin{equation}\label{jointprior}
f\left(s_{1},\ldots,s_{N}\vert \mathop{\mathbf{X}_{1}}^{P\times T},\ldots,\mathbf{X}_{N}\right) \propto \alpha^{M-1}\mathop{\prod}_{m=1}^{M}g(\mathbf{X}^{\ast}_{m})(n_{m} -1)!,
\end{equation}
where our notation conditions on the $\left(\mathbf{X}_{\ell}\right)$ for emphasis, though this prior doesn't treat them as random.  In our mixture formulations, we define
\begin{equation}
g(\mathbf{X}^{\ast}_{m}) = \int\mathop{\prod}_{\ell:s_{\ell} = m}f\left(\mathbf{X}_{\ell}\vert \mathbf{\Delta}_{\ell},\mathbf{H}_{x}\right)f\left(\mathbf{\Delta}_{\ell}\vert\mathbf{\Theta}^{\ast}_{x,m}\right)f\left(\mathbf{\Theta}^{\ast}_{x,m}\right)
d\mathbf{\Delta}_{\ell}d\mathbf{\Theta}^{\ast}_{x,m}d\mathbf{H}_{x},
\end{equation}
with $\mathbf{\Theta}^{\ast}_{x,m} = \{\mathbf{\Lambda}^{\ast}_{x,m}, \tau^{\ast}_{x,m}, \rho^{\ast}_{x,m}\}$. The form of $g(\mathbf{X}^{\ast}_{m})$ slightly generalizes \cite{Muller:2011:PPM} from a DP to our DP mixture.  \cite{Muller:2011:PPM} highlight that it is not necessary for ``similarity" function, $g(\mathbf{X}^{\ast}_{m})$, to be specified as random.  It should be invariant to predictor labels and their scale, and assign larger probabilities of co-clustering where $\left(\mathbf{X}_{\ell}\right)_{\ell: s_{\ell} = m}$ are closer in value. We use a symmetric random probability distribution which possesses these properties for computational convenience.

The formulation of Equation~\ref{jointprior} is also equivalent to replacing the single random distribution, $G$, with a collection, $\left(G_{x} = \mathop{\sum}_{h=1}^{\infty} p_{xh}\delta_{\{\mathbf{\Lambda}^{\ast}_{y,h}, \bm{\kappa}^{\ast}_{h}\}}\right)$, that indexes weights, $(p_{xh})$, by the predictor values (such that, marginally, each $G_{x}$ is a DP). Counties with similar predictor values are assigned a relatively higher prior probability of co-clustering.
\section{Posterior Computation} \label{mcmc}
We implement the posterior computations for the predictor-indexed mixture model, specified in Section~\ref{predassist} (from which it is easy to derive the computations for the mixture model of Section~\ref{nopred}), in a sequential scan of parameter blocks from their full conditional posterior distributions in the \texttt{growfunctions2} package for \textbf{R}\citep{R}, which is written in \texttt{C++} for fast computation and available from the authors on request.  We briefly highlight aspects of our posterior sampling algorithm for the major sets of parameters, below:
\begin{enumerate}
\item Model for $\mathbf{Y} = (y_{bq})$
    \begin{enumerate}
    \item Sample each block of $P\times T$ random effect coefficients, $\left(\mathbf{B}_{\ell}\right)$, independently, using the elliptical slice sampler (ESS) of \citet{murray2010} for block-sampling parameters under a multivariate Gaussian prior (that we generalized to matrix variate Gaussian distributions).  The ESS generates ($P\times T$) proposals through a convex combination of a draw from the prior and the previously sampled value. The proposals lie on the ellipse parameterized with a phase angle. The ESS uses a slice sampling algorithm \citep{Neal00slicesampling} to draw proposals for the phase angle. Proposals are evaluated with the likelihood,
        \begin{equation} \label{likeB}
        L\left(\mathbf{B}_{\ell}\right) = \mathop{\prod}_{b \in b(\ell)}\mathop{\prod}_{q \in q(b)}f\left(\tilde{y}_{bq,\ell}\vert \mathop{\sum}_{j \in q}\mathbf{x}_{\ell j}^{'}\bm{\beta}_{\ell j},\sigma_{bq}^{2}\right),
        \end{equation}
        where $b(\ell)$ denotes the (usually multiple) blocks in which county $\ell$ is nested.  Similarly, $q(b)$, denotes the often multiple periods, $q$, linked to block, $b$. We define $\tilde{y}_{bq,\ell} = y_{bq} - \mathop{\sum}_{\ell^{'} \neq \ell \in b}\mathop{\sum}_{j \in q}\mathbf{x}_{\ell^{'}j}^{'}\mathbf{\beta}_{\ell^{'}j}$ to subtract out estimated functions for all other counties, $\left(\ell^{'}\right) \neq \ell$, which are also linked to $y_{bq}$.
    \item Sample the posterior distribution for locations of the GP covariance in by-cluster groups, $\left(\kappa^{\ast}_{dm}\right)_{d = 1,\ldots,D}$, from the subset of counties, ($\mathbf{B}_{\ell}$), assigned to that cluster because $\kappa^{\ast}_{dm} \independent \kappa^{\ast}_{dm^{'}}$ for $m^{'} \neq m$, \emph{a posteriori}, in a Metropolis-Hastings scheme using the following log-posterior kernel,
        \begin{eqnarray}
        &&\mbox{log}f\left(\kappa^{\ast}_{dm}|\bm{\kappa}^{\ast}_{-dm},\mathbf{s},
        \mathbf{\Lambda}^{\ast}_{y,m},
        \{\mathbf{B}_{\ell}:s_{\ell} = m\}\right) \nonumber\\
         &&\propto -\frac{1}{2}n_{m}P~\mbox{log}\left(|\mathbf{C}\left(\kappa_{dm}^{\ast}\right)|\right)
        -\frac{1}{2}\mbox{tr}\left[\sum_{\ell:s_{\ell} = m}\mathbf{C}\left(\kappa_{dm}^{\ast}\right)\mathbf{B}_{\ell}^{'}
        \mathbf{\Lambda}^{\ast}_{y,m}\mathbf{B}_{\ell}\right] \label{posttheta}\\
        &&+ (a-1)\log(\kappa^{\ast}_{dm}) - b\kappa^{\ast}_{dm} \nonumber,
        \end{eqnarray}
        where  $(a,b)$ are shape and rate hyperparameters of a gamma prior, respectively, which are both set equal to $1$.  This posterior representation is a relatively straightforward Gaussian kernel of a non-conjugate probability model.

        We adapt a Metropolis-Hastings algorithm of \citet{wang:2013} for sampling \emph{each} $\kappa^{\ast}_{dm}$ that is designed to speed computation  by introducing a lower-dimensional temporary space where the likelihood (e.g. the $T \times T$, Gaussian process covariance matrix, $\mathbf{C}$) is approximated using a subset of the $T$ time-points.   We develop a transition / proposal distribution based on composing moves in the lower dimensional, temporary space (using a slice sampler), where computations of the lower-dimensional GP covariance matrix are fast.  If the lower dimensional approximations are relatively good, this approach will speed chain convergence by producing draws of lower autocorrelation since each proposal includes a sequence of moves generated in the temporary space for drawing an equivalent effective sample size.  See \citet{savitskygp:2014} for more details.
    \item Sample location, $\mathbf{\Lambda}^{\ast}_{y,m}$, from a $P$ dimensional Wishart posterior with degrees of freedom, $n_mT+(P+1)$ and $P\times P$ inverse scale, $\mathop{\sum}_{\ell:s_{\ell} = m}\mathbf{B}_{\ell}\mathbf{C}(\bm{\kappa}^{\ast}_{m})
        \mathbf{B}_{\ell}^{'} + \mathbb{I}_{P}$.
    \item Sample cluster assignments, $\mathbf{s} = \left(s_{1},\ldots,s_{N}\right)$, from their full conditionals using the P\'{o}lya urn representation, \citet{blackwell:1973},
        \begin{equation}\label{posts}
            f\left(s_{\ell}=s| \mathbf{s}_{-i},\mathbf{\Theta}^{\ast}_{s},\alpha,
            \tau_{\epsilon},\mathbf{B}_{\ell},\mathbf{\Delta}_{\ell}\right) \propto
            \begin{cases}
            \frac{n_{-\ell,s}}{n-1+\alpha}L\left(\mathbf{B}_{\ell},\mathbf{\Delta}_{\ell}\right) & \text{if $1 \leq s  \leq M^{-}$}\\
            \frac{\alpha/c^{*}}{n-1+\alpha}L\left(\mathbf{B}_{\ell},\mathbf{\Delta}_{\ell}\right) & \text{if $s = M^{-} + h$},
            \end{cases}
        \end{equation}
        where $n_{-\ell,s} = \sum_{\ell^{'} \neq \ell}\mathbb{I}(s(\ell^{'}) = s)$ is the number of counties, excluding unit $\ell$, assigned to cluster $s$, so that units are assigned to an existing cluster with probability proportional to its ``popularity" and $M^{-}$ denotes the total number of clusters when unit $\ell$ is removed (which is equal to $M$ unless $\ell$ is a member of singleton cluster).  The posterior assigns a county (through $s_{\ell}$) to a new cluster with probability proportional to $\alpha d_{0}= \int \mathcal{N}\left(\mathbf{B}|\bm{\kappa},\ldots\right)G_{0}(d\bm{\kappa})$, that requires the likelihood to be integrable in closed form with respect to the base distribution, which is not the case under our non-conjugate parameterization through the GP covariance matrix.  So we utilize the auxiliary Gibbs sampler formulation of \citet{neal:2000} and sample $c^{\ast} \in \mathbb{N}$ (typically set equal to $2$ or $3$) locations from base distribution, $G_{0}$, \emph{ahead} of any assigned  observations, to define $h = M^{-} + c^{\ast}$ candidate clusters in an augmented space.  We then draw $s_{\ell}$ from this augmented space, where any location not assigned units (over a set of draws for $\mathbf{s}$) is dropped.
    \end{enumerate}
\item Model for $\mathbf{X}_{\ell} = (\mathbf{x}_{\ell j})$
    \begin{enumerate}
    \item Sample $P\times T,~\mathbf{\Delta}_{\ell}$, independently, by stacking the transpose of the $P~,T\times 1$ rows of $\mathbf{\Delta}_{\ell}$ to form the $PT \times 1,~\bm{\delta}_{v,\ell}$, from which we perform a draw from the following conjugate Gaussian posterior,
        \begin{equation}
        f\left(\bm{\delta}_{v,\ell}|\mathbf{X}_{\ell},\mathbf{H}_{x},\mathbf{s},
        \mathbf{\Lambda}^{\ast}_{x,s_{\ell}},\mathbf{Q}\left(\tau^{\ast}_{x,s_{\ell}},
        \rho^{\ast}_{x,s_{\ell}}\right)\right) = \mathcal{N}_{PT}\left(\mathbf{h}_{\delta},\bm{\phi}_{\delta}^{-1}\right),
        \end{equation}
        where we define $PT \times 1,~\mathbf{e}_{\delta} = \mathbf{H}_{x,T}\mathbf{x}_{v,\ell}$, with $\mathbf{H}_{x,T} = (\mathbf{H}_{x} \otimes \mathbb{I}_{T})$, while $\mathbf{x}_{v,\ell}$ is formed by stacking the transpose of the rows of $\mathbf{X}_{\ell}$.  Posterior precision, $\bm{\phi}_{\delta} = \mathbf{H}_{x,T} + \left(\mathbf{\Lambda}^{\ast}_{x,s_{\ell}} \otimes \mathbf{Q}\left(\tau^{\ast}_{x,s_{\ell}},\rho^{\ast}_{x,s_{\ell}}\right)\right)$. Finally, compose $\mathbf{h}_{\delta} = \bm{\phi}_{\delta}^{-1}\mathbf{e}_{\delta}$.
    \item Sample the location parameters, $\left(\tau^{\ast}_{x,m}\right)$, of the $T \times T$ CAR precision matrix, $\mathbf{Q}$, from the Gamma distribution,
        \begin{equation}
        f\left(\tau^{\ast}_{x,m}\vert (\mathbf{\Delta}_{\ell}:s_{\ell} = m),\rho^{\ast}_{x,m}\right) = \mathcal{G}a\left(a_{1},b_{1}\right),
        \end{equation}
        with shape, $a_{1} = 0.5n_{m}TP + a$, and rate, $b_{1} = 0.5\mbox{tr}\left[\mathop{\sum}_{\ell:s_(\ell) = m} \mathbf{R}^{\ast}_{m}\mathbf{\Delta}_{\ell}^{'}\mathbf{\Lambda}^{\ast}_{x,m}
        \mathbf{\Delta}_{\ell} + b\right]$, where $\mathbf{R}^{\ast}_{m} = \left(\mathbf{D}_{x} - \rho^{\ast}_{m}\mathbf{\Omega}_{x}\right)$.

        Next, sample $\rho^{\ast}_{x,m}$ using a slice sampler with the following posterior evaluation kernel,
        \begin{equation}
        \begin{split}
        \quad\log f\left(\rho^{\ast}_{x,m}\vert (\mathbf{\Delta}_{\ell}:s_{\ell} = m),\tau^{\ast}_{x,m}\right) \\
        \quad \propto 0.5n_{m}P\log\vert R^{\ast}_{m}\vert + 0.5\tau^{\ast}_{x,m}\rho^{\ast}_{x,m}\mbox{tr}\left[\mathop{\sum}_{\ell:s_{\ell} = m}\mathbf{\Omega}_{x}\mathbf{\Delta}_{\ell}^{'}\mathbf{\Lambda}^{\ast}_{x,m}
        \mathbf{\Delta}_{\ell}\right].
        \end{split}
        \end{equation}
    \end{enumerate}
\end{enumerate}

\section{Results for the ACS} \label{acs}
\begin{figure}[!h]
\begin{center}
\includegraphics[width=4.0in,height=2.7in]{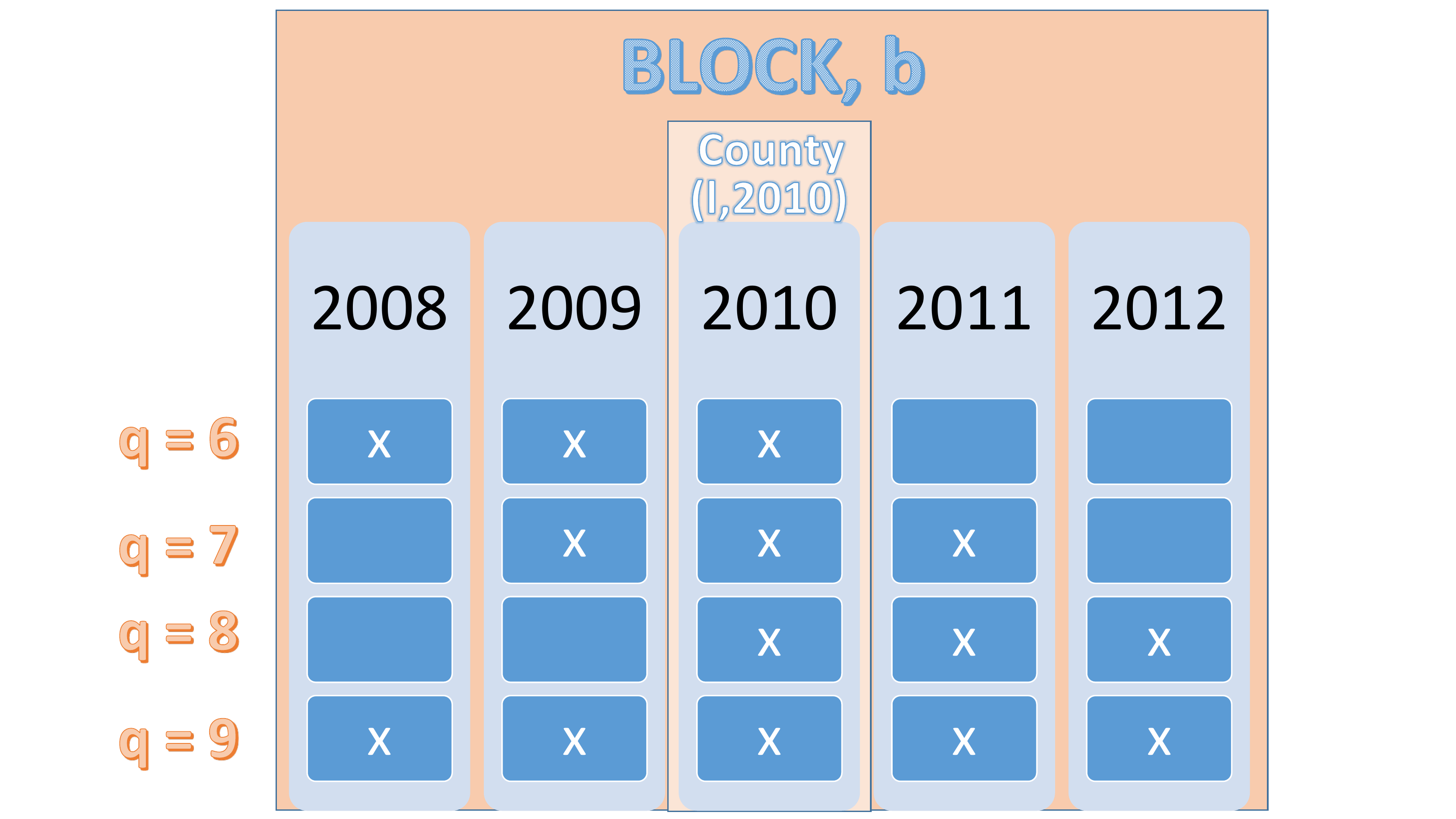}
\caption{Conceptual Illustration of Multiple Data Points Linked to each County.}
\label{datalink}
\end{center}
\end{figure}

Our likelihood of Equation~\ref{modellike} sums the county-year parameters, $\left(f_{\ell j}\right)$, nested in each block-period statistic, $y_{bq}$.  Conversely, there are multiple statistics (indexed by block-period) that link to each county-year parameter, which provide some information to support the estimation of the that parameter.  Figure~\ref{datalink} presents a conceptual illustration for a hypothetical county, ``$\ell$", linked to a block, ``$b$", where block $b$, in turn, includes published observations for $3-$ and $5-$ year periods.  Each row of Figure~\ref{datalink} indicates with an ``$x$", the link of the associated period to the five years of time points in our ACS dataset.  Suppose we are interested to recover the associated statistics linked to the $\ell-2010$ county-year for block $b$.  The highlighted column for $2010$ reveals there are five statistics for block $b$ that nest $\ell-2010$ and provide some information for its estimation.  There will potentially be many observed statistics used to estimate $\ell-2010$ in the case it nests in multiple blocks.

We next illustrate estimation results by comparing the fitted function for a selected county with the collection of statistics to which it is linked at each time point. To make the comparison meaningful, we only want to include the portion of each statistic that provides information about that county; for example, if a county is nested, along with other counties, in a metropolitan area for which we have an observed statistic, $y_{bq}$, we'd like to extract from the statistic only the portion of the observed employment level that provides information about that county.  We compute a ``pseudo" statistic, $\tilde{y}_{bq,\ell j}$, in Equation~\ref{pseudopoint} for each block, $b$, and period, $q$, linked to a latent, county-year function parameter, $f_{\ell j} = \mathbf{x}_{\ell j}^{'}\bm{\beta}_{\ell j}$, by subtracting away from statistic, $y_{bq}$, (to which county-year, $\ell-j$, is linked) all other estimated county-year function values  (besides that for $\ell-j$) for which $y_{bq}$ also provides information (including years $(j^{\ast})$ other than $j$ for county $\ell$). The quantity $\hat{\bm{\beta}}_{\ell^{\ast} j^{\ast}}$ in Equation~\ref{pseudopoint} represents the posterior mean of the sampled values from our MCMC.  (Of course, coefficient values are sampled at each iteration of the MCMC under Equation~\ref{modellike} for estimation. So we could rao-blackwellize over the posterior draws for the coefficient values to create a pseudo statistic, but it is less intuitive than our proposed construction in Equation~\ref{pseudopoint}).
\begin{equation} \label{pseudopoint}
\tilde{y}_{bq,\ell j} = y_{bq} - \mathop{\sum}_{\ell^{\ast} \neq \ell \in b}\mathop{\sum}_{j \in q} \mathbf{x}^{'}_{\ell^{\ast} j}\hat{\bm{\beta}}_{\ell^{\ast} j} - \mathop{\sum}_{j^{\ast} \neq j \in q} \mathbf{x}^{'}_{\ell j^{\ast}}\hat{\bm{\beta}}_{\ell j^{\ast}}
\end{equation}

Equation~\ref{likeB} demonstrates that the posterior for each matrix of $P\times T$ coefficients, $\mathbf{B}_{\ell}$, weights the contribution of each statistic, $y_{bq}$, in proportion to its precision (inverse variance), such that statistics associated to block-periods closer in geography (that nests relatively fewer counties) and time exert more influence on the estimated result.  Our presentation of results, to follow, will illustrate the fit mechanism by plotting each pseudo-statistic, $\tilde{y}_{bq,\ell j}$ for county-year, $\ell-j$, with size of the displayed point in proportion to its precision.

The next set of figures illustrate estimated functions for selected counties as compared to the associated pseudo statistics under the DP mixtures of Gaussian processes model of Section~\ref{nopred}.  We subsequently compare the fit performances for the clustering prior formulations of Sections~\ref{predassist} and \ref{nopred}, which include and exclude predictors, respectively, in the prior for cluster assignments.

\begin{figure}[!h]
\begin{center}
\includegraphics[width=3.8in,height=3.0in]{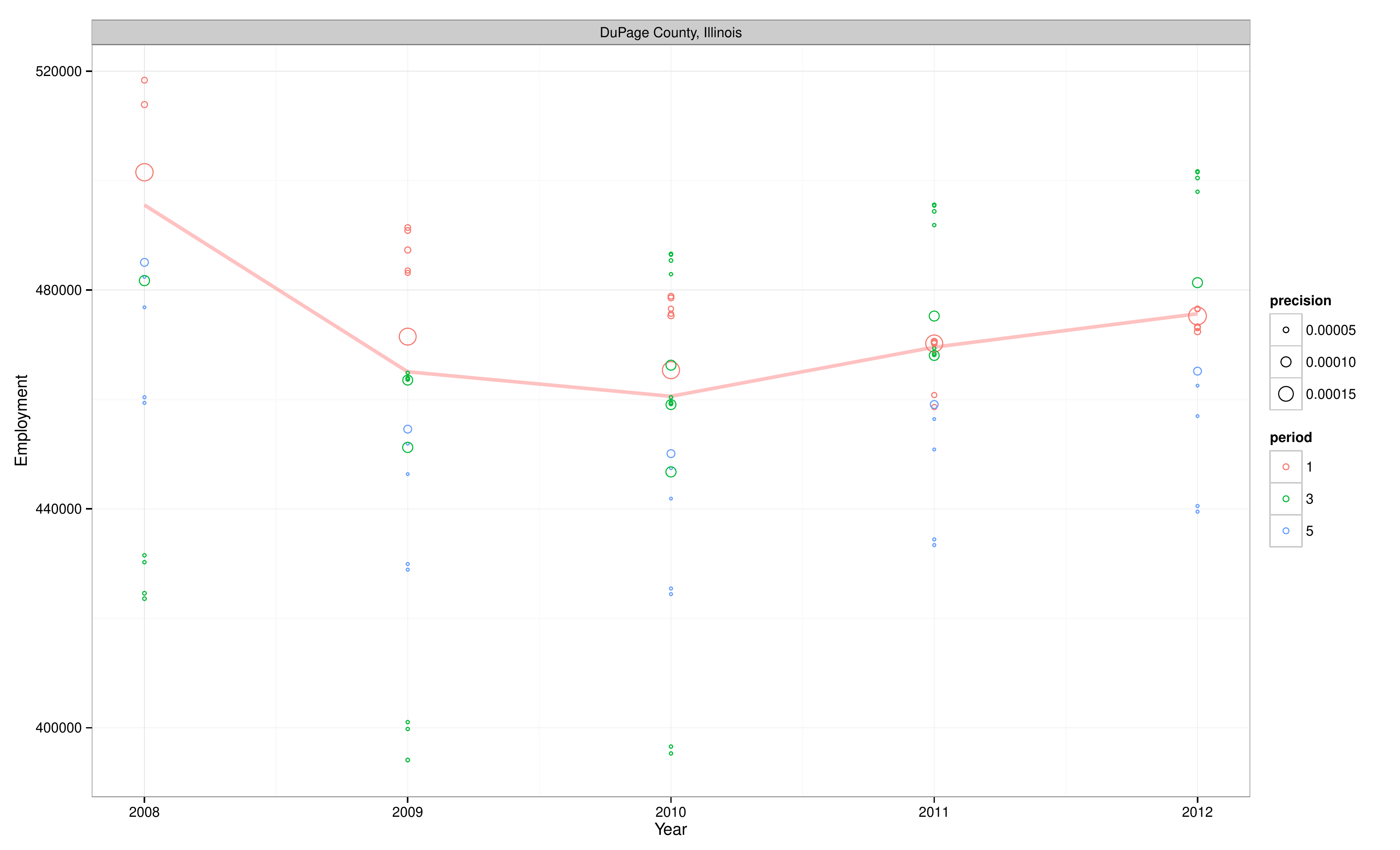}
\caption{Estimated Function vs. Pseudo Data for $1-$ year county: Fitted function (pink line) compared to the collection of pseudo data points in each year, $2008-2012$, for a large-sized (by population) county, DuPage County, IL, with published $1-$ year period estimates. Each hollow circle represents a pseudo statistic and its size is proportional to its estimated precision.  Each hollow circle is colored based on the period of the data point; red denotes a $1-$ year period, green denotes a $3-$ year period and red denotes a $5-$ year period. }
\label{acs1yr}
\end{center}
\end{figure}

Figure~\ref{acs1yr} displays the fitted function (in the pink line), along with the collections of pseudo statistics in each year for a county with $1-$ year period ACS observations.  The size of each pseudo statistic is in proportion to its precision, with $1-$ year period points colored in red, $3-$ year period points in green and $5-$ year period points in blue.  Since this county has observed $1-$ year period statistics, those will be the most precise (and, hence, largest) for estimating this county.  Nevertheless, we see that while the fitted trend is similar to that expressed by the $1-$ year estimates, it differs because the fitted values are influenced by pseudo statistics representing other blocks in which this county nests. These blocks provide additional information about employment levels for the county.  While the fitted values are more influenced by pseudo statistics that express higher precision, they are also influenced by the number of such points around a given value.  Here, we see a good coherence between the sets of $3-$ and $5-$ year period estimated pseudo statistics for blocks nesting this county in $2009-2011$ (time points $2-4$).  These values lie below the $1-$ year values and pull down the fitted function away from the $1-$ year period estimate.

We may \emph{not} use these pseudo data plots to assess the fit quality, however, precisely because of the pseudo statistics are convolved with the estimation procedure.  We may, nevertheless, comment on the coherence or closeness among estimated pseudo statistics with relatively larger precision values, which offers comment on the strength of estimation.

\begin{figure}[!h]
\begin{center}
\includegraphics[width=3.8in,height=3.0in]{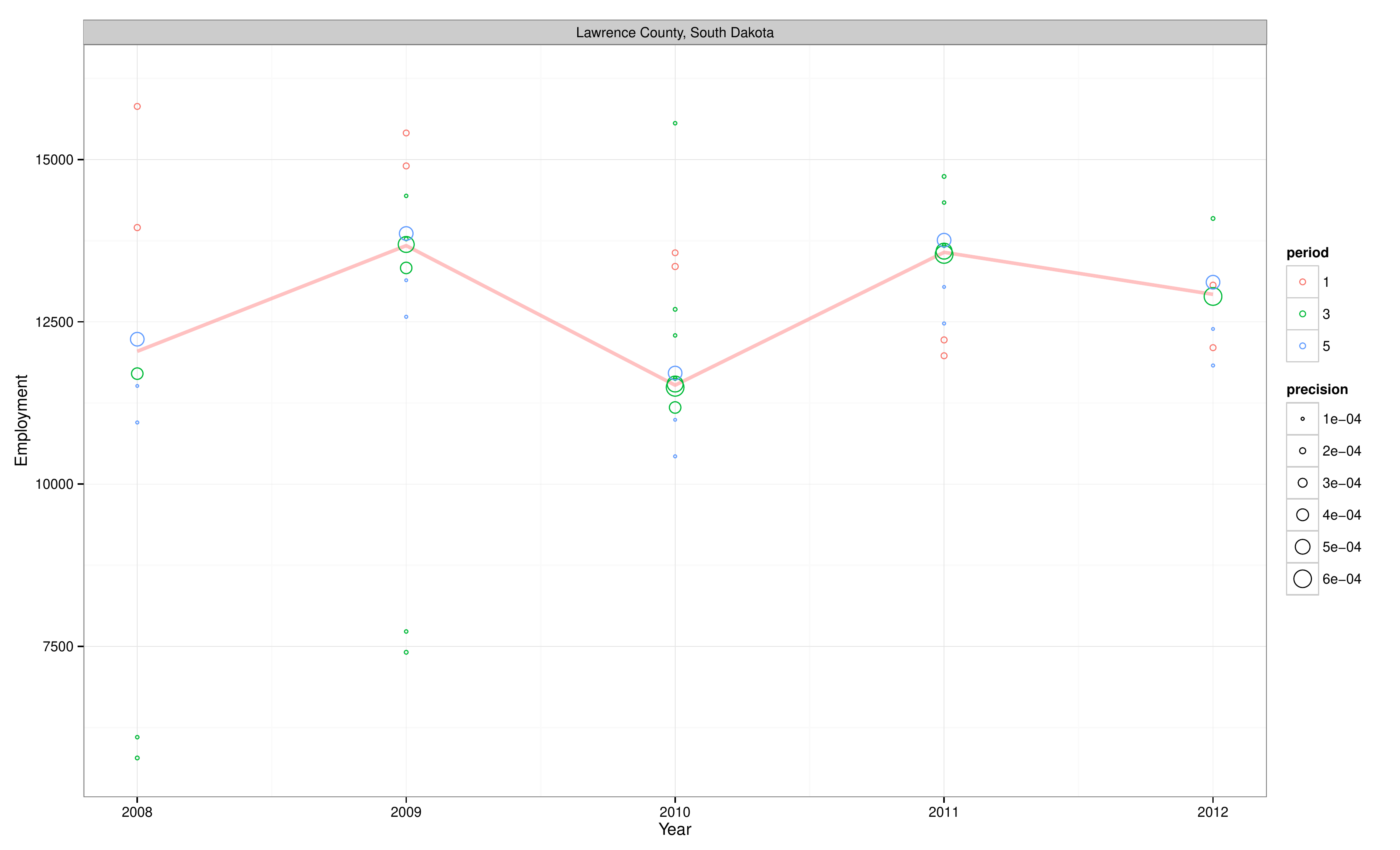}
\caption{Estimated Function vs. Pseudo Data for $3-$ year county: Fitted function (pink line) compared to the collection of pseudo data points in each year, $2008-2012$, for a medium-sized (by population) county, Lawrence County, SD, with published $3-$ year (but not $1-$ year) period estimates. Each hollow circle represents a pseudo statistic and its size is proportional to its estimated precision.  Each hollow circle is colored based on the period of the data point. }
\label{acs3yr}
\end{center}
\end{figure}

\begin{figure}[!h]
\begin{center}
\includegraphics[width=3.8in,height=3.0in]{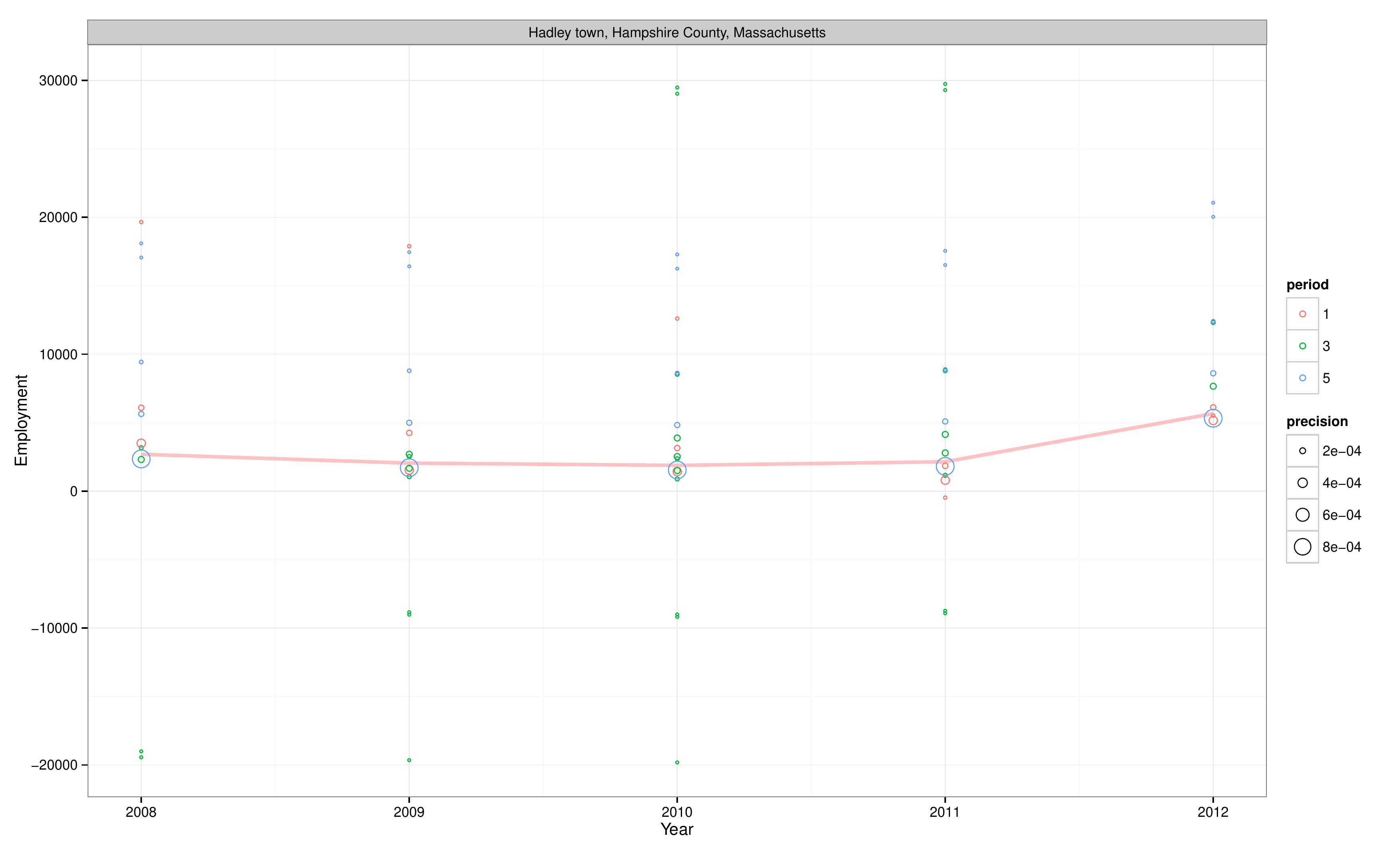}
\caption{Estimated Function vs. Pseudo Data for $5-$ year county: Fitted function (pink line) compared to the collection of pseudo statistics
in each year, $2008-2012$, for a small-sized (by population) township (MCD), Hadley, Hampshire County, MA, with published $5-$ year (but not $1-$ or $3-$ year) period estimates. Each hollow circle represents a pseudo statistic and its size is proportional to its estimated precision.  Each hollow circle is colored based on the period of the data point. }
\label{acs5yrnear2}
\end{center}
\end{figure}

Figure~\ref{acs3yr} displays the estimated function compared to pseudo statistics for a county with $3-$ and $5-$ year period observations, but not $1-$ period observations.  We see a good coherence between estimated pseudo statistics among near in size blocks in which this county nests.
Figure~\ref{acs5yrnear2} presents an MCD for which only a single $5-$ year period estimate is available.  The results also express a good coherence between the relatively higher precision pseudo statistics because every New England MCD nests in a county, which in this case also has $1-$ year period statistics.  

We observe in these figures that some of the pseudo statistics are very large in magnitude - highly positive or negative - though their small precisions result in their exerting little-to-no influence in the estimation of the functions.
The overly high magnitude values occur where a county is nested in an area far different in size than itself; e.g. nested in a balance of metropolitan areas, which will potentially include hundreds of counties.  While a state-level estimate may be relatively precise for estimating a large, state-level quantity, it is highly imprecise for estimating a small, constituent piece.  Thus, there is almost no information borrowed from a block that is far larger in size than a constituent county, reflecting a limitation in the ability of the model to borrow information.

In general, we find that the QCEW super sector employment level predictors helps to identify the county-year functions by providing magnitude information and regulating the shrinkage of by-county regression coefficients where the county employment levels span vast differences in size of their populations and labor markets.  Yet, the resulting modeled estimate is typically quite different in level and trend (not shown) than the total of the QCEW super sector employment values.  We are not surprised because the QCEW provides place-of-work employment from establishments, while the ACS is a household survey providing place-of-residence employment.

\begin{figure}[!h]
\begin{center}
\includegraphics[width=3.5in,height=2.8in]{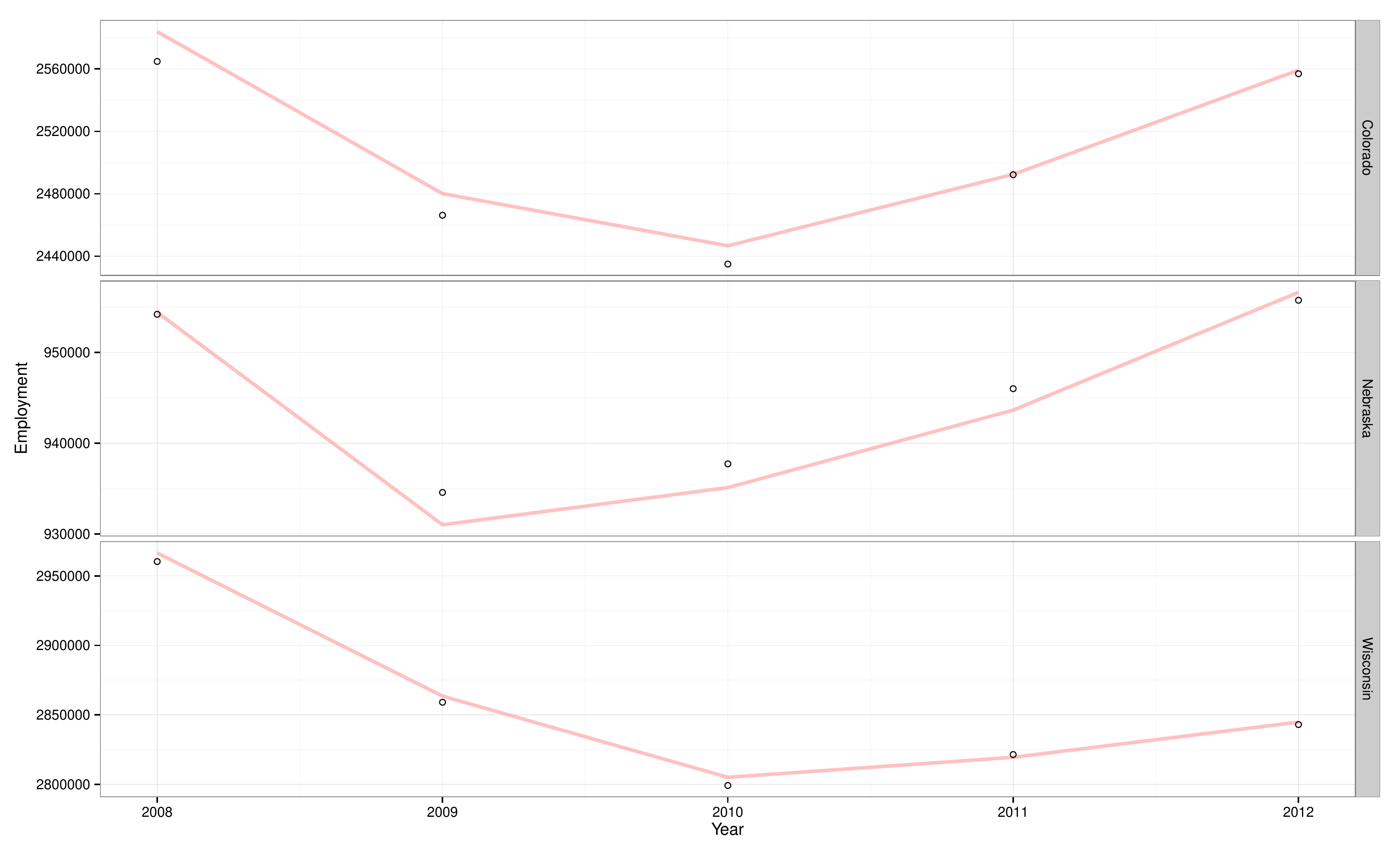}
\caption{County-year fitted values summed to state-level (pink line) versus data values (hollow circles) for randomly-selected states. }
\label{worststates}
\end{center}
\end{figure}

Our estimation model entirely focuses on estimating fine-level, county-year parameters, using blocks and periods that nest them.  Nevertheless, we've seen that there is limited information provided to estimate county by a block observation nesting it which is much larger (in population and employment) than the county.  So, since $74\%$ of counties lack $1-$ year period estimates, a question arises about the quality of estimation at the state level composed by summing over the county-year parameters nested in each state-year. The roll-up of estimated functions to the states produces estimates for all states that are within $1-2\%$ of $1-$ year period state-level estimates in the ACS.
Figure~\ref{worststates} shows the estimated summed functions compared to the observed data points for three randomly-selected states, which illustrates the estimation of latent functions at the county-year level provides a good estimation for state-level, $1-$ year period observations.

\subsection{Assessment of Fit Quality}
\begin{figure}[!h]
\begin{center}
\includegraphics[width=5.5in,height=3.8in]{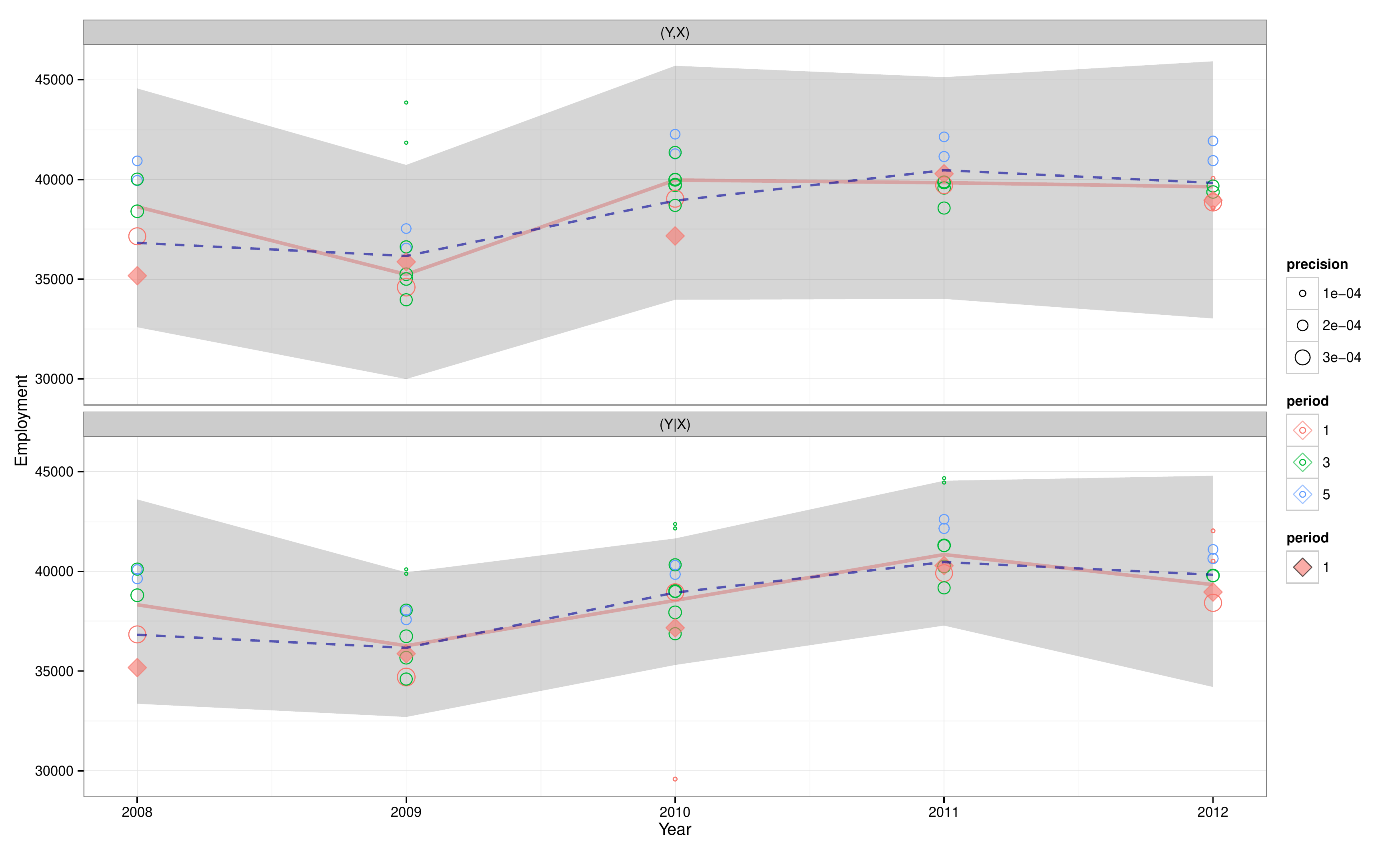}
\caption{Comparison of model-estimated values for a $1-$ year county (Craven County, NC) when excluding $1-$ year
data values.  The top plot panel provides results for the predictor-assisted clustering model (which we label, $(Y,X)$),
while the plot in the bottom panel excludes predictors in the prior for cluster assignments (which we label, $(Y|X)$).  The solid, pink line in each plot panel presents the posterior mean fitted function when excluding the $1-$ year data points, while the dashed, blue line presents the posterior mean when including the $1-$ data points.  The gray shading represents the $95\%$ credible intervals as estimated on the models excluding $1-$ year data points.  The associated pseudo statistics are also estimated from the models excluding $1-$ year data points.  The solid pink diamonds plot the $1-$ year data points.}
\label{craven}
\end{center}
\end{figure}

We may not directly assess the fit performance of the estimated county-year functions for $3-$ and $5-$ year counties due to the absence of observed $1-$ year data values.  An indication of fit quality may, however, be provided by \emph{excluding} the (five) $1-$ year data values for a county \emph{with} available $1-$ year data values and comparing how the models - that exclude or include predictors in the prior distributions for cluster assignments - estimate the county-year function to when the $1-$ year values are included. Figure~\ref{craven} presents estimated county-year functions for Craven County, North Carolina.  The top panel displays estimated results under the predictor-assisted clustering model of Section~\ref{predassist}, while the bottom panel displays the same under the model that excludes predictors (in the prior for assignment to clusters) of Section~\ref{nopred}.  The solid, pink line in each plot panel presents the posterior mean fitted function when excluding the $1-$ year data values, while the dashed, blue line presents the same when including the $1-$ year values.  The gray shading displays the associated $95\%$ credible intervals under exclusion of the $1-$ year data values and the associated pseudo statistics are also constructed using the fitted functions under exclusion of these values.  Finally, the pink, diamond points display the $1-$ year data values.

We explored a number of $1-$ year counties, at random, and found a high-degree of similarity between the estimated county-year functions with and without inclusion of the $1-$ year data values under both models.  The model excluding predictors in the prior for the cluster assignments of Section~\ref{nopred}, however, tends to consistently express slightly less difference in estimated functions with and without inclusion of the $1-$ year data values.  We present Craven County as something of a worst-case result that provides clearer differentiation between the performances of the two models.  Craven County is a relatively small, $1-$ year county.  The Craven County $1-$ year data points would suggest increasing employment through the Great Recession period of $2008-2010$, which is antipodal to most counties in North Carolina (and the U.S., as a whole).  The estimated employment trend when \emph{including} the $1-$ year data values, which is displayed in the dashed blue line, actually estimates an employment decline from $2008-2009$, followed by a recovery in $2009-2010$.  The other blocks (in addition to the county, itself) that include Craven County favor a decline - recovery trend, as may be observed in the associated co-plotted pseudo statistics.  The Craven County estimation scheme balances the (higher precision, more reliable) $1-$ year data values with the information conveyed by the blocks at multiple resolutions in which Craven County nests.

We see that both models amplify the estimated employment decline from $2008-2009$ when the $1-$ year data values are \emph{excluded}, which effectively increases the influence of the other blocks containing Craven County.   Yet, the model excluding predictors in assigning clusters well-captures both the increasing trend from $2010-2011$ and the decreasing trend from $2011-2012$.  The predictor-assisted clustering model expresses a slightly steeper decline, followed by a more rapid recovery.  It is likely the case that our predictors, which intend to measure the composition of the economic activity of a county, induced co-clustering among counties with this pattern during the Great Recession.  The fitted results under both models may be sensitive to the composition of the county-year predictors because they are below the resolution of the observed data; for example, perhaps if we include additional predictors that provide information about poverty concentration or education achievement the predictor-assisted model may or may not out-perform.  In any case, given the estimation sensitivity to predictor values, they should be carefully chosen based on their ability to comment on the economic conditions of each county.  These results generally suggest that the spatial and temporal nesting construction that underpin our models may provide reasonable estimates across counties.  The larger credible intervals for the predictor-assisted clustering model reflects the large space of partitions or clusterings induced when including the predictors in the prior for the mixing measure.

Table~\ref{tab:fit} provides fit statistics for the models including ($(Y,X)$) and excluding predictors ($(Y|X)$) in the estimation of clusters.  We display the $DIC_{3}$ criterion \citep{Cele:Forb:Robe:Titt:repl:2006} that focuses on the marginal (predictive) density $\widehat{f\left(\mathbf{y}\right)}$ in lieu of $f(\mathbf{y}|\widehat{\mbox{parameters}})$, which is more appropriate for mixture models.  Also shown is the log-pseudo marginal likelihood that employs ``leave-one-out" cross-validation \citep{Gelf:Dey:baye:1994}. We estimate $\prod_{r=1}^{BQ}f\left(y_{r}|\mathbf{y}_{-r},M_{k}\right)$, (where $r$ denotes a block-period case observation), the $\log$ of which is the log pseudo marginal likelihood (LPML), where $M_{k}$ indexes a model.  We employ a weighted re-sampling of parameters from existing posterior draws in a fashion that provides model parameter samples from $f\left(\mbox{parameters}|\mathbf{y}_{-r},M_{k}\right)$ \citep{Ster:Cres:post:2000}.  This approach reduces the known sensitivity to outliers expressed by the LPML.  Our primary modeling goal, however, is not ``out-prediction", beyond the data, but ``in-prediction" at a resolution lower than the observed data.  We, nevertheless, see that the predictor-assisted clustering model doesn't provide a notably better mean deviance, $\bar{D}$, than the simpler model to justify the added complexity.  The similar fit statistics, combined with the lower perturbation in the estimated functions illustrated in Figure~\ref{craven}, incline us to prefer the simpler model of Section~\ref{nopred}.
\begin{table}[!h]
\centering
\begin{tabular}{rrr}
  \hline
  \hline
 & $(Y,X)$ & $(Y|X)$ \\
  \hline
-LPML & $233517$ &   $228181$  \\
  DIC$_{3}$ & $449663$ &  $450199$  \\
  $\bar{D}$ & $444634$ &  $446928$  \\
   \hline
\end{tabular}
\caption{Fit performance comparison between model including predictors in prior for cluster assignments,$(Y,X)$ ,and model excluding predictors in clustering, $(Y|X)$. Lower values indicate better fit performance for all included fit statistics.}
\end{table}\label{tab:fit}

\section{Simulation Study} \label{sim}
\begin{figure}[!h]
\begin{center}
\includegraphics[width=3.5in,height=3.0in]{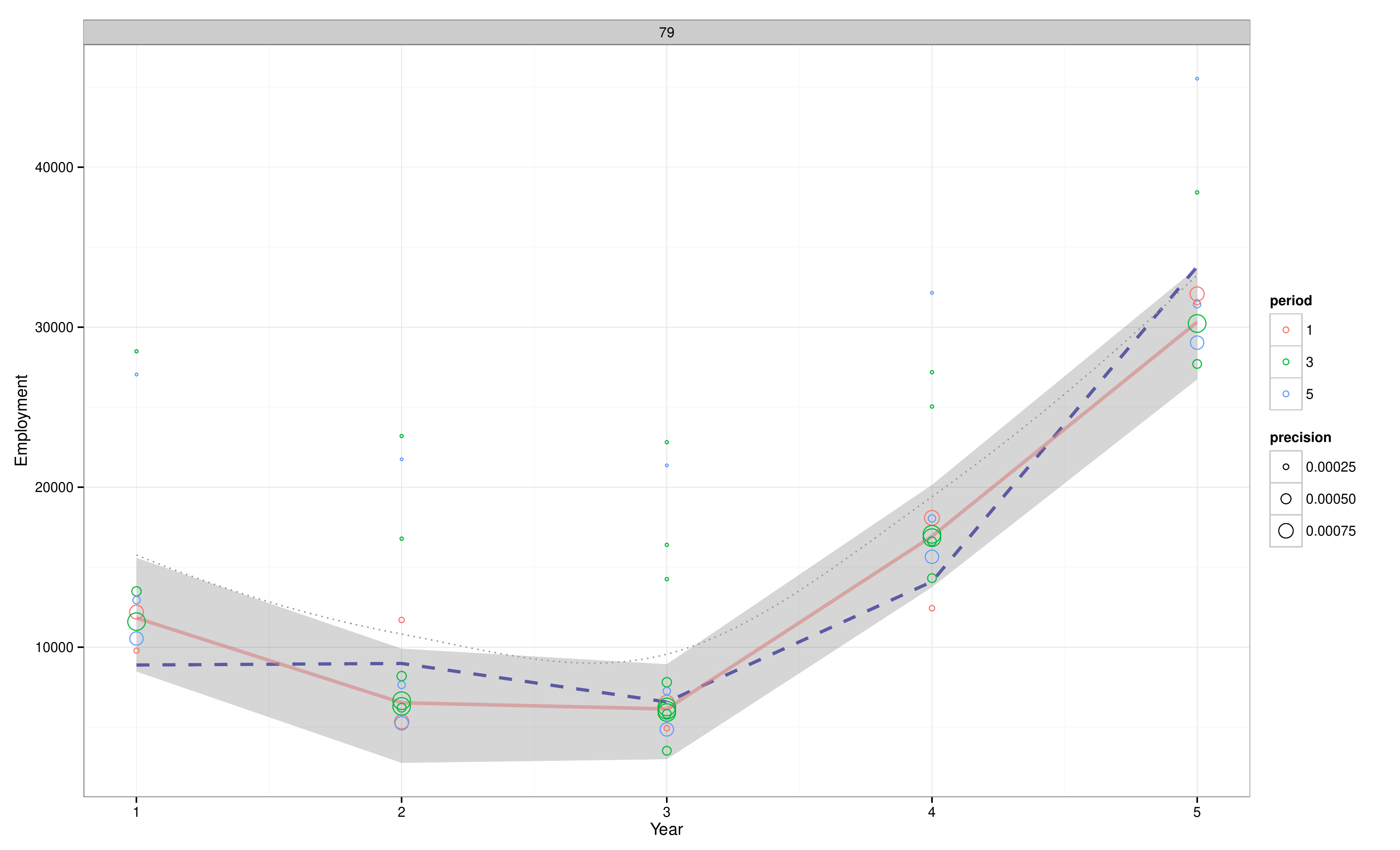}
\caption{Fitted versus data values for simulated $3-$ year county.}
\label{sim3yr}
\end{center}
\end{figure}

Our examination of results for the ACS helped provide insight on the fit performance, but perhaps does not fully address the quality of fit for counties with only $3-$ and $5-$ year data values.  To address quality of fit for these counties, we generate synthetic values for coefficients, $\left(\mathbf{B}_{\ell}\right)$, from Equation~\ref{matprior}, employing the posterior means of covariance parameters $\left(\hat{\mathbf{\Lambda}}_{y,\ell},\hat{\bm{\kappa}}_{\ell}\right)$ from the model of Section~\ref{nopred}.  We next compute $f_{\ell j} = \mathbf{x}_{\ell j}^{'}\beta_{\ell j}$, where $\mathbf{X}_{\ell}$ is observed (known). We next generate $y_{bq} \ind \mathcal{N}\left(\mathop{\sum}_{\ell \in b}\mathop{\sum}_{j \in q}f_{\ell j},\sigma_{bq}^{2}\right)$.
The same nesting relationships of (county, year) to (block, period) from the ACS are duplicated for the simulation study, so that we are generating a synthetic version of ACS employment counts.  Of course, this simulation assumes that our spatial and temporal nesting construction is the correct generating model, which we do not know to the case, though the fit performances on $1-$ year counties when excluding the $1-$ year data values suggests that this assumption may be broadly reasonable.
Figure~\ref{sim3yr} presents the pseudo statistics, fitted function (denoted by a pink line) and associated $95\%$ credible interval (denoted by gray shading), along with the true function (denoted by the dashed, blue line) for a $3-$ year county. It reveals that our model also does well on a county for which we have $3-$ year period statistics, but not $1-$ year period statistics.

\begin{figure}[!h]
\begin{center}
\includegraphics[width=3.5in,height=3.0in]{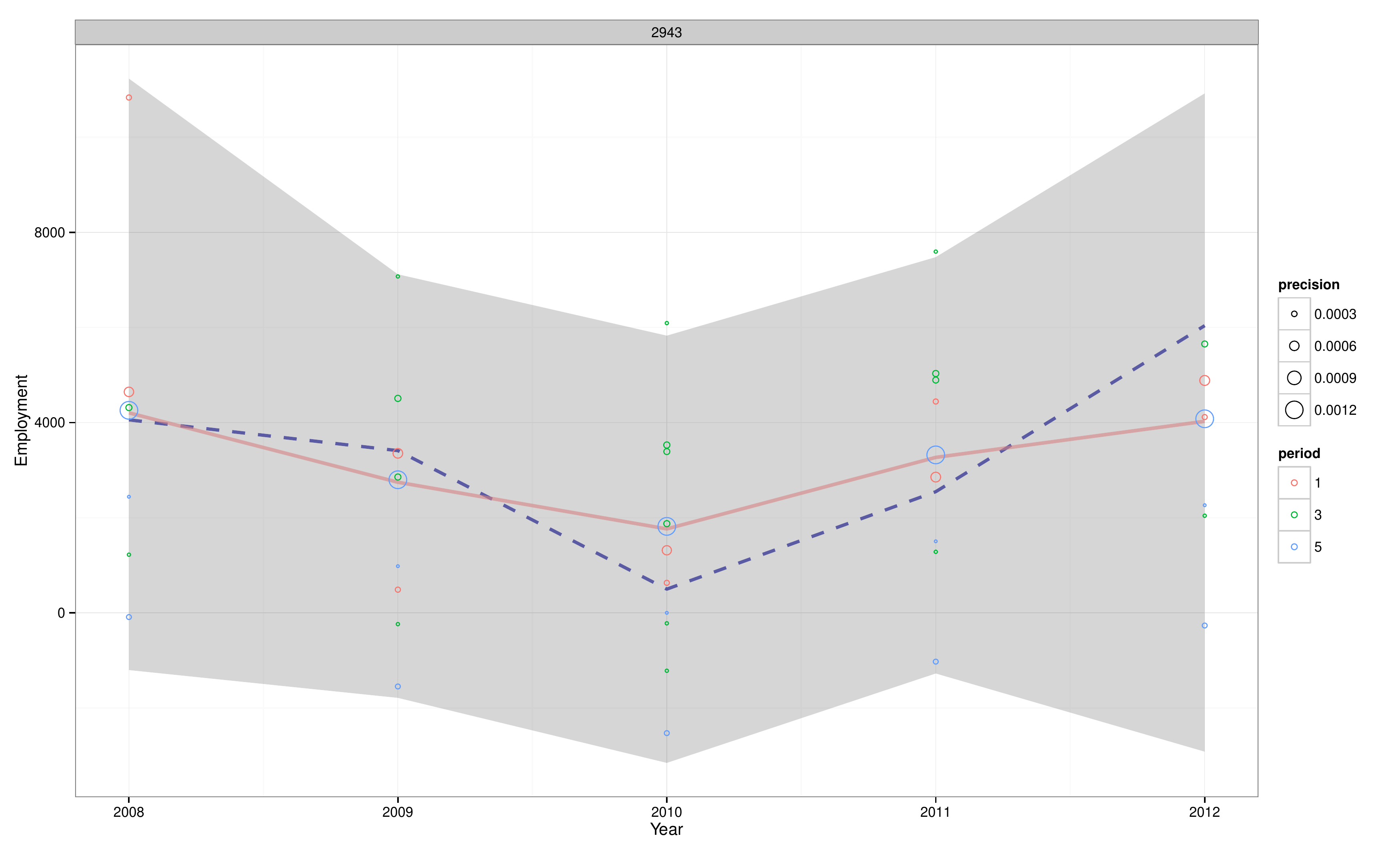}
\caption{Fitted versus Data values for simulated $5-$ year county linked to one or more blocks of similar size.}
\label{sim5yrnear2}
\end{center}
\end{figure}

Similarly to the $3-$ year county result, Figure~\ref{sim5yrnear2} presents typical results for a county with only a single, $5-$ year statistic available in the case where that county is nested in a block relatively near to it in size.  As earlier mentioned, this situation is typical for MCD's, which by construction (in New England) are nested within counties.  While we see that the fitted result expresses more smoothness than the truth, it does generally follow local features in the true trend and the credible interval is wider than those for counties with published $3-$ year period statistics.

\begin{figure}[!h]
\begin{center}
\includegraphics[width=3.5in,height=3.0in]{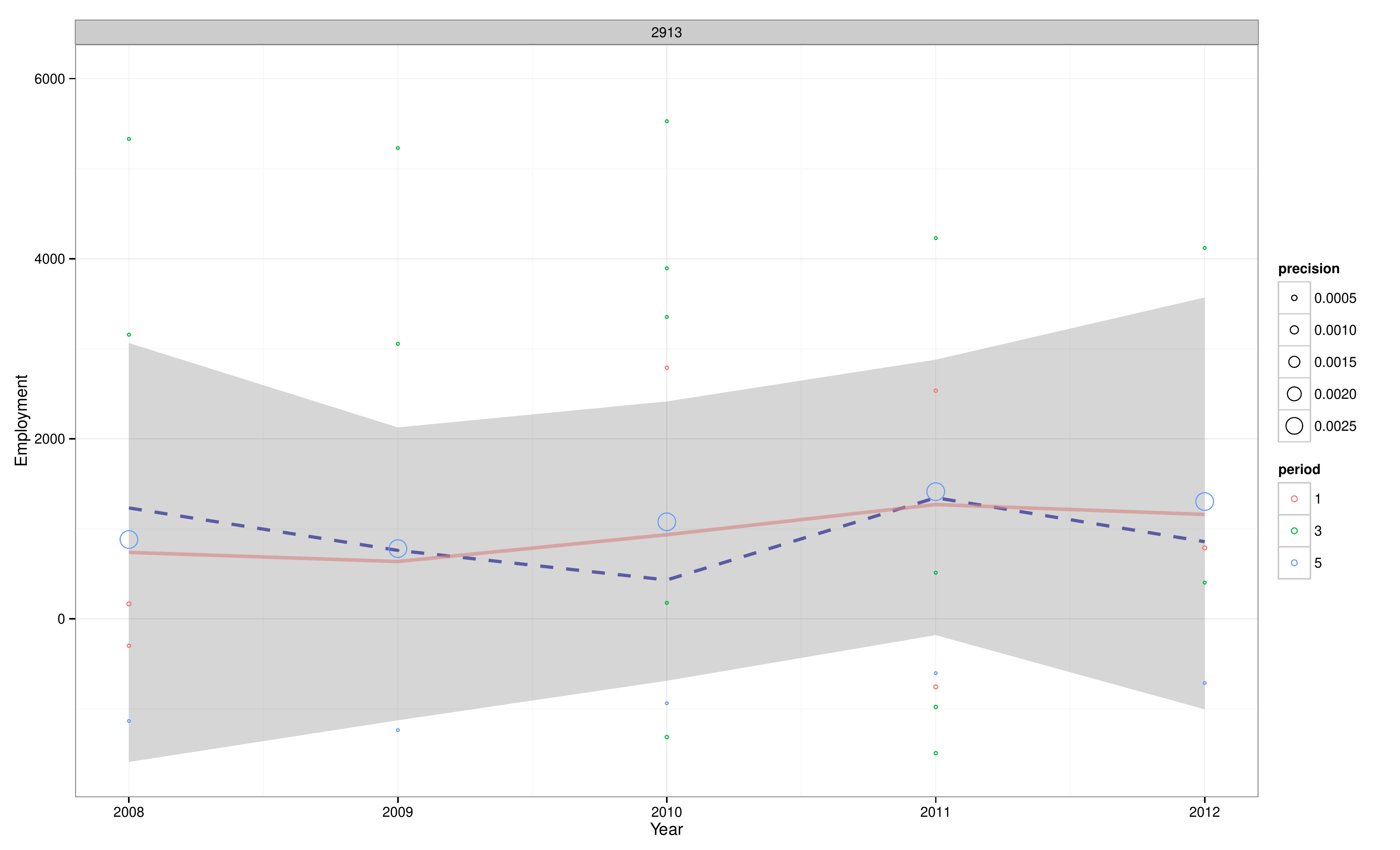}
\caption{Fitted versus data values for simulated $5-$ year county linked only to blocks much larger in size.}
\label{sim5yrfar}
\end{center}
\end{figure}

Figure~\ref{sim5yrfar} presents estimated results for a county with only a single $5-$ year period observed statistic and that is nested in a block far different (much larger) in size.  The true trend is similar to that in Figure~\ref{sim5yrnear2} and we see that the fitted function expresses a greater degree of over-smoothing and is unable to capture local features in time, though the overall true trend and magnitude are still captured.  Adding data for upcoming years will bring in additional $5-$ year period statistics, which are expected to improve the quality of estimation for these far-nested counties by borrowing strength over periods, rather than blocks.

\section{Discussion} \label{discussion}
Motivated by the use of ACS employment data at the BLS to allocate statewide CPS employment estimates to sub-state, local areas, we have developed a general approach to estimate fine-scale time and areal-indexed parameters using an ensemble of coarse-scale observations that spatially and temporally nest the parameters.  We specify the likelihood to link subsets of the parameters that exhaustively nest each block-period observation.  Our best-performing Bayesian multiscale model of Section~\ref{nopred} formulates a relatively simple nonparametric mixture model for estimating the latent county functions in a fashion that facilitates the shrinking together of similar functions by the data.  The flexible shrinking under the Bayesian non-parametric approach, which penalizes complexity, combined with leveraging nesting relationships to identify an ensemble of observations that provide information about each latent parameter, provides a broadly useful approach.

Many ACS users, such as the LAUS program in BLS, would prefer to employ $1-$ year period statistics for counties, but are relegated to using $5-$ year period published statistics in the case where analyses are conducted across all counties in the U.S.  Results from our simulation study demonstrate that our approach performs well to uncover the latent true county-year parameters for $3-$ year counties and $5-$ year counties, where the $5-$ year counties nest within similarly-sized blocks (along with few other counties).  There was some notable over-smoothing of the estimated county function (though the magnitude and global trend are captured) for $5-$ year counties exclusively nested in much larger-sized blocks, which occurs because we only have a single, $5-$ year period statistic for these counties.  We expect improvements in the fit accuracy for these counties as we add upcoming years to the five years of data that we considered for our analysis because our mixtures of Gaussian process formulations borrows strength across years.  Employing an ensemble of statistics published at varied resolutions even adds value for the estimation of counties with $1-$ year period statistics by incorporating the additional statistics associated to blocks nesting each $1-$ year county. Our approach may be applied to any variable from the ACS, as well as to other data sets that express this multiresolution structure.

\section{Acknowlegdements}
The authors wish to thank our colleagues at the Bureau of Labor Statistics whose focus on continuous improvement led them to sponsor this project.  We thank the following important contributors:
\begin{enumerate}
\item Sean B. Wilson, Senior Economist, who led the effort to gain approval for the project and evaluate the quality of results.
\item Garrett T. Schmitt, Senior Economist, who helped us structure the data and associated county-block links.
\end{enumerate}

\bibliography{mv_refs_nov2014}
\bibliographystyle{agsm}


\end{document}